\documentclass[showpacs,showkeys,preprintnumbers,superscriptaddress,amsmath,floatfix,amssymb,a4paper,secnumarabic]{revtex4}
\usepackage[colorlinks=true]{hyperref}
\usepackage{graphicx}
\usepackage{epsfig}
\usepackage{slashbox}

\newcommand{\beq}{\begin{equation}}
\newcommand{\eeq}{\end{equation}}
\newcommand{\bal}{\begin{aligned}}
\newcommand{\eal}{\end{aligned}}

\def\tr{\hbox{Tr}}
\def\dalam{\hbox
{\vrule\vbox{\hrule\hbox to 1ex{ \hfill}\kern 1 ex\hrule}\vrule}}
\def\sign{\hbox{sign}}
\def\1/2{\hbox{$ {1 \over 2}$ }}

\def\h{\hbar}
\def\i/h{{i \over \h}}

\def\arctg{\hbox{arctg}}

\def\inf{\infty}
 
\def\v{\vec}

\def\a{\alpha} 
\def\b{\beta} 
 
\def\g{\gamma} \def\G{\Gamma} 
\def\d{\delta} \def\D{\Delta}
\def\l{\lambda} \def\L{\Lambda}
\def\e{\epsilon} \def\E{\hbox{$\cal E $}}

\def\s{\sigma}
\def\r{\rho}

\def\vf{\varphi} 
\def\f{\phi} \def\F{\Phi}
 
\def\p{\psi} \def\P{\Psi}

\def\m{\mu}
\def\n{\nu}

\def\k{\kappa}  \def\vk{\varkappa}

\def\z{\zeta}

\def\tt{\theta}

\def\<{\langle}
\def\>{\rangle}

\def\({\left(}
\def\[{\left[}
\def\){\right)}
\def\]{\right]}

\begin{document}

\title{Non-perturbative vacuum polarization effects in two-dimensional supercritical Dirac-Coulomb system. II. Vacuum energy}

\author{A.~Davydov}
\email{davydov.andrey@physics.msu.ru} \affiliation{Department of Physics and
Institute of Theoretical Problems of MicroWorld, Moscow State
University, 119991, Leninsky Gory, Moscow, Russia}

\author{K.~Sveshnikov}
\email{costa@bog.msu.ru} \affiliation{Department of Physics and
Institute of Theoretical Problems of MicroWorld, Moscow State
University, 119991, Leninsky Gory, Moscow, Russia}

\author{Y.~Voronina}
\email{voroninayu@physics.msu.ru} \affiliation{Department of Physics and
Institute of Theoretical Problems of MicroWorld, Moscow State
University, 119991, Leninsky Gory, Moscow, Russia}

\begin{abstract}
Non-perturbative vacuum polarization effects   are explored for  a supercritical Dirac-Coulomb system with $Z > Z_{cr,1}$ in 2+1 D, based on the original combination of analytical methods, computer algebra and numerical calculations, proposed recently in Refs.~\cite{davydov2017}\nocite{sveshnikov2017}-\cite{voronina2017}. Both the  vacuum charge density $\r_{VP}(\v r)$ and vacuum energy $\E_{VP}$ are considered. Due to a lot of details of calculation the whole work is divided into two parts I and II.  Taking account of results, obtained in the part I ~\cite{partI2017} for $\r_{VP}$, in  the present part II the evaluation of the vacuum energy $\E_{VP}$ is investigated with emphasis on the   renormalization and convergence of the partial expansion for $\E_{VP}$. It is shown that the  renormalization via fermionic loop turns out to be the universal tool, which removes the divergence of the theory both in the purely perturbative and essentially non-perturbative regimes of the vacuum polarization. The main result of calculation is that for a wide range of the system parameters in the overcritical region $\E_{VP}$  turns out  to be a rapidly decreasing function  $\sim - \eta_{eff}\, Z^3/R\, $ with $\eta_{eff}>0 $ and $R$ being the size of the external Coulomb source. To the end the similarity in calculations of  $\E_{VP}$ in 2+1 and 3+1 D is discussed, and qualitative arguments are presented in favor of the possibility for complete screening of the classical electrostatic energy of the Coulomb source by the vacuum polarization effects for  $Z \gg Z_{cr,1}$ in 3+1 D.


\end{abstract}

\maketitle

\section{Introduction}

This paper continues the work, initiated in Refs.~\cite{davydov2017}\nocite{sveshnikov2017}-\cite{voronina2017} and devoted to the study of non-perturbative QED-effects, caused by diving of discrete levels into the lower continuum in supercritical static or adiabatically slowly varying Coulomb fields, that are created by localized extended sources with $Z> Z_{cr}$.   Such effects have  attracted a considerable amount of theoretical and experimental activity (see Refs.~\cite{raf2017}\nocite{NP2015, popov2015, superheavy2013}-\cite{ruffini2010} and refs. therein), since in 3+1  QED for $Z > Z_{cr,1} \simeq 170$ a non-perturbative reconstruction of the vacuum state is predicted, which should be accompanied by a number of nontrivial effects including the vacuum positron emission (see Refs.~\cite{raf2017},\cite{rg1977}\nocite{greiner1985,plunien1986}-\cite{greiner2012} and refs. therein). Similar in essence effects are expected to come out also both in  2+1 D (planar graphene-based hetero-structures \cite{katsnelson2006,shytov2007,nomura2007,kotov2008,pereira2008,herbut2010,wang2013,nishida2014}) and  in 1+1 D (one-dimensional ``hydrogen ion''\cite{barb1971,krainov1973,shabad2006,shabad20061,shabad2008,oraevskii1977,karn2003,vys2014}).

Recently, in Refs.~\cite{davydov2017}\nocite{sveshnikov2017}-\cite{voronina2017} an original combination of analytic methods, computer algebra tools and numerical calculations has shown, that for a wide range  of the system parameters in the one-dimensional Dirac-Coulomb model the nonlinear effects  could lead in the supercritical region to the behavior of the vacuum energy, substantially different from the perturbative quadratic growth  up to (almost) quadratic decrease into the negative region $\sim -|\eta| Z^2$. In the present work, which consists of two parts  I and II,  these methods are applied to the study of  similar vacuum polarization effects  for a  2+1  Dirac-Coulomb system in the overcritical region. More concretely, in the part I ~\cite{partI2017} we have considered the vacuum charge density $\r_{VP}$, while in  the present part II by taking into account the results, obtained in I, the behavior of the vacuum energy $\E_{VP}$ is explored.

As in the part I, the external Coulomb field is chosen in the form of a projection onto a plane of the potential of the uniformly charged sphere with radius  $R$
\beq \label{1.00}
A^{ext}_{0}(\v r)=Z |e| \[\frac{1}{R}\tt\(R-r\)+\frac{1}{r}\tt\(r-R\) \] \ ,
\eeq
leading to the potential energy
\beq
\label{1.1}
V(r)= - Z \a \[\frac{1}{R}\tt\(R-r\)+\frac{1}{r}\tt\(r-R\) \] \ .
\eeq
Compared to the model of the uniformly charged ball this potential is more preferable, since it allows to perform the most part of calculations in the analytical form, while the evaluation of critical charges shows that  in both cases the final answers should be quite close. And although with the standard choice of the fine-structure coupling $\a \simeq 1/137$ and without special selection of the Coulomb field cut-off parameters such a system could be treated only as a toy-model of the 3+1 D problem,  its study for $Z > Z_{cr}$ should be of considerable interest, since it allows to reproduce almost all the properties of the realistic 3+1 D problem of vacuum polarization by superheavy nuclei or nuclear quasi-molecule, but with certain substantial simplifications caused by the smaller number of rotational degrees of freedom.  For these reasons the radius of the external source  is chosen as in 3+1 D for the case of superheavy nuclei
\beq
\label{1.2}
R=R(Z) \simeq 1.2\, (2.5\, Z)^{1/3} \ \text{fm} \ .
\eeq

It should be specially noted that we do not consider here the question of the origin and self-energy of the Coulomb sources generating the potential (\ref{1.00}). Namely, within the purely planar problem such sources cannot be treated even as the localized ones, since their charge density should decrease for $r \to \inf$ as $\sim 1/r^3$. So it is indeed the potential  (\ref{1.00}), that is primary in the present work, and only the vacuum polarization effects caused by it are  explored,   not the field self-energy of those planar sources that formally match the potential (\ref{1.00}) within the purely two-dimensional problem. So in our  work the vacuum polarization energy,
 that is generated by the quantized electron-positron field, can be found, whereas the classical one, on the contrary, cannot, because it is necessary to know how and from what such a planar system was created. The latter problem statement could be actual for graphene, but in this case it is necessary to consider the regime of strong coupling with $\a_g \sim 1$ and perform a substantial refinement of the Coulomb field cut-offs,  due to which the whole  picture of vacuum effects in the supercritical region undergoes global changes, concerning first of all the numerical aspects. Therefore the study of such effects in the planar Dirac-Coulomb system with parameters similar to  graphene on the substrate will be considered separately.

As in other works on vacuum polarization in the strong  Coulomb field, radiative corrections from virtual photons are neglected. Henceforth, if it is not stipulated separately, relativistic units  $\hbar=m_e=c=1$ are used. Thence the coupling constant $\a=e^2$ is also dimensionless, what significantly simplifies the subsequent analysis, while the numerical calculations, illustrating the general picture, are performed for  $\a=1/137.036$.

\section{Perturbation Theory for the Vacuum Energy in 2+1 QED}

In the  2+1 QED the vacuum polarization energy to the first order of the perturbation theory (PT) is given by the following expression
\beq
\label{2.1}
\E^{(1)}_{VP}=\frac{1}{2} \int d^2r\, \r^{(1)}_{VP}(\vec{r})A_{0}^{ext}(\vec{r}) \ ,
\eeq
where $\r^{(1)}_{VP}(\vec{r})$ is the first-order vacuum charge density, which is found via the corresponding vacuum polarization (Uehling) potential
\beq
\label{2.2}
\r^{(1)}_{VP}(\vec{r})=-\frac{1}{4 \pi} \D_{2}\, A^{(1)}_{VP,0}(\vec{r}) \ , \eeq
where  $\D_2$ is the two-dimensional Laplace operator.
In its turn, the Uehling potential $A^{(1)}_{VP,0}$ is expressed by means of the polarization operator $\Pi_R(-\vec{q}\,^2)$ and the Fouriet-transform of the external potential $\widetilde{A}_{0}(\vec{q})$ \cite{greiner2012}
\beq
\label{2.3}
\bal
&A^{(1)}_{VP,0}(\vec{r})=\frac{1}{(2 \pi)^2} \int d^2q\,\mathrm{e}^{i \vec{q} \vec{r}} \Pi_{R}(-q^2)\widetilde{A}_{0}(\vec{q}) \ , \\
&\widetilde{A}_{0}(\vec{q})=\int d^2 r'\,\mathrm{e}^{-i \vec{q} \vec{r\,}' }A^{ext}_{0}(\vec{r}\,' ) \ , \qquad q=|\vec{q}| \ ,
\eal
\eeq
where
\beq
\label{2.4}
\Pi_R(-q^2)=\frac{\a}{2q} \[\frac{2 }{q} + \(1-\frac{4}{q^2}\) \arctg \( \frac{q}{2}\)\] \ .
\eeq

From (\ref{2.3}) and (\ref{2.4}) for the external field (\ref{1.00}) one obtains the expression for the Uehling potential in the form of an axial-symmetric function (for details of calculation see part I, Appendix A)
\beq
\label{2.5}
\bal
A_{VP,0}^{(1)}(r)&=\frac{Z \a |e|}{4}\int\limits_{0}^{\infty}dq\,\frac{J_0(qr)}{q}\[\frac{2 }{q} + \(1-\frac{4}{q^2}\) \arctg \( \frac{q}{2}\)\] \\
&\times \(2\[1+J_1(q R)-q R J_0(q R)\] +\pi q R\[ J_0(q R) \mathbf{H}_1(q R)- J_1(q R) \mathbf{H}_0(q R)\]\) \ ,
\eal
\eeq
with  $J_{\n}(z)$ and  $\mathbf{H}_{\n}(z)$ being the Bessel and Struve functions correspondingly. The vacuum density, determined from (\ref{2.2}) and (\ref{2.5})
\beq
\label{2.5c}
\bal
\r_{VP}^{(1)}(r)&=\frac{Z \a |e|}{16\pi}\int\limits_{0}^{\infty}dq\,q J_0(q r)\[\frac{2 }{q} + \(1-\frac{4}{q^2}\) \arctg \( \frac{q}{2}\)\] \\
&\times \(2\[1+J_1(q R)-q R J_0(q R)\] +\pi q R\[ J_0(q R) \mathbf{H}_1(q R)- J_1(q R) \mathbf{H}_0(q R)\]\)
\eal
\eeq
is finite for all $r\neq R$ with logarithmic singularity for $r \to  R$.

 In the next step, from  (\ref{2.1}) and (\ref{2.5c}) one finds the vacuum polarization energy to the first order of PT
\beq
\label{2.6}
\bal
\E^{(1)}_{VP}&=\frac{(Z \a)^2}{32}\int\limits_{0}^{\infty}dq\,\[\frac{2 }{q} + \(1-\frac{4}{q^2}\) \arctg \( \frac{q}{2}\)\] \\
&\times \(2\[1+J_1(q R)-q R J_0(q R)\] +\pi q R\[ J_0(q R) \mathbf{H}_1(q R)- J_1(q R) \mathbf{H}_0(q R)\]\)^2.
\eal
\eeq

Let us also mention  that within  PT the integral vacuum charge in  the linear approximation  vanishes exactly
\beq
\label{2.7}
\int \! d^2r \ \r^{(1)}_{VP}(r)= 0 \ .
\eeq
Moreover, in the case under consideration the direct check, performed in the part I,  shows that upon renormalization the vacuum charge $Q_{VP}^{ren}=\int \! d^2r \ \r_{VP}^{ren}(r)$  turns out to be  non-vanishing only for $Z>Z_{cr,1}$ due to non-perturbative effects, caused by diving of discrete levels into the lower continuum in accordance with  Refs.~\cite{raf2017},\cite{rg1977}-\cite{greiner2012}. And in what follows it will be shown, how the latter circumstance shows up in the behavior of the vacuum energy in the overcritical region.

\section{Vacuum Polarization Energy for $Z>Z_{cr,1}$: General Properties}

The starting expression for the vacuum energy $ \E_{VP} $ is given by
\beq
\label{f39}
\E_{VP}=\<H_D\>_{vac}=\1/2 \(\sum\limits_{\e_n<\e_F} \e_n-\sum\limits_{\e_n \geq \e_F}\e_n\) \ ,
\eeq
where $\e_F$ is the Fermi level, which in such problems with the external Coulomb field should be chosen at the threshold of the lower continuum ($\e_F=-1$), while $\e_{n}$ are the energy eigenvalues of the corresponding spectral Dirac-Coulomb problem (DC)
\beq
\label{f391}
\(- i \,\vec{\a}\,\vec{\nabla}+V(\vec{r})+\b -\e_n \)\p_n(\vec{r})=0 \ .
\eeq
As it was shown in the part I, in 2+1 D the degeneracy of each energy eigenstate with fixed $m_j$ equals to  2, and in what follows this factor will be explicitly shown in all the expressions for $\E_{VP}$, while the spectral DC problem  without any loss of generality will be considered in the two-dimensional representation with  $\a_i=\s_i$, $\b=\s_3$. The only exception is the lowest discrete level in each partial channel with fixed $m_j$, whose degeneracy turns out to be twice less. This circumstance is discussed in detail below (see formulae (\ref{sum_d})-(\ref{sum_d1})).

The expression  (\ref{f39}) is obtained from the Dirac hamiltonian, written in the form that is invariant under charge conjugation, and is defined up to a constant, depending on the choice of the energy origin \cite{rg1977, greiner1985, plunien1986, greiner2012, raf2017}. It follows from (\ref{f39}) that $\E_{VP}$ is negative and divergent even in absence of external fields $ A_{ext} = 0 $. But since the vacuum charge density $\r_{VP}$ is defined so that  it vanishes identically for $A_{ext} = 0 $ (see part I, expr.(11)), the natural choice for the normalization of $ \E_{VP} $ should be the same. Besides this, in the presence of the external Coulomb potential of the type (\ref{1.00}) there appears in the sum (\ref{f39}) also an (infinite) set of discrete levels. To pick out  exclusively the interaction effects it is therefore necessary to subtract from each discrete level the mass of the free electron at rest.

Thus, in the physically well-motivated form and in agreement with $\r_{VP}$, the initial expression for the vacuum energy should be written as
\beq
\label{f40}
\E_{VP}=\1/2 \(\sum\limits_{\e_n<\e_F} \e_n-\sum\limits_{\e_n \geq \e_F} \e_n + \sum\limits_{-1\leq \e_n<1} \! 1
\)_A \ - \ \1/2 \(\sum\limits_{\e_n<0} \e_n-\sum\limits_{\e_n
	>0} \e_n \)_0 \ ,
\eeq
where the label A denotes the  non-vanishing external field $A_{ext}$, while the label 0 corresponds to the free case with $A_{ext}=0$. The vacuum energy, defined in such a way, vanishes by turning off the external field, while by turning on it contains only the interaction effects, hence, the expansion of $ \E_{VP} $ in (even) powers of the external field (\ref{1.00}) should start from $ \mathrm{O}(Z^2) $.

Now let us extract  from (\ref{f40}) separately the contributions from the discrete and continuous spectra for each fixed $ m_j $, and afterwards use for the difference of integrals over the continuous spectrum $ (\int\mathrm{d}k\sqrt{k^2+1})_A-(\int\mathrm{d}k\sqrt{k^2+1})_0 $ the well-known technique, which represents this difference in the form of an integral of the elastic scattering phase $ \d(k) $. Such techniques have been quite effectively applied to evaluation of the one-loop quantum corrections to the soliton mass in essentially nonlinear QFT models in 1+1 D (see Refs.~\cite{raja1982},\cite{sv1991} and refs. therein), and in Refs.~\cite{davydov2017}-\cite{voronina2017} to the vacuum energy calculation in the  1+1-dimensional DC problem with the external potential of the type (\ref{1.00}). Omitting a number of almost obvious steps of computation, that have been considered in detail in Ref.~\cite{sveshnikov2017}, let us  write the final answer
\beq
\label{f41a}
\E_{VP}=2\sum\limits_{m_j=1/2,3/2,..} \E_{VP,|m_j|} \nonumber \eeq \beq\label{f41} =2\sum\limits_{m_j=1/2,3/2,..} \({1 \over 2\pi} \int\limits_0^\inf \!   \  \frac{k \, \mathrm{d}k }{\sqrt{k^2+1}} \ \d_{tot,|m_j|}(k) + {1 \over 2} \sum\limits_{-1 \leq \e_{n, \pm m_j}<1} \(2-\e_{n,+m_j}-\e_{n,-m_j}\)\) \ .
\eeq
In (\ref{f41a}) $ \d_{tot,|m_j|}(k) $ is the total phase shift for the given values of the wavenumber $k$ and modulus of the total momentum $|m_j|$, including the  contributions from the scattering states from both  continua and  $\pm m_j$ for the two-dimensional radial DC problem
\beq
\label{f41b}
\left\lbrace\bal
&\frac{d}{d r}\p_1(r)+\frac{1/2-m_j}{r}\,\p_1(r)=(\e-V(r)+1)\p_2(r) \ ,\\
&\frac{d}{d r}\p_2(r)+\frac{1/2+m_j}{r}\,\p_2(r)=-(\e-V(r)-1)\p_1(r) \ ,
\eal\right.
\eeq
to which the spectral problem (\ref{f391})  for the axial-symmetric potential of the type (\ref{1.00}) is reduced by means of the substitution
\beq
\p(\vec{r})={1 \over \sqrt{2 \pi} }\, \begin{pmatrix}
i \p_{1}(r)\,\mathrm{e}^{i(m_j-1/2)\vf}\\
\p_{2}(r)\,\mathrm{e}^{i(m_j+1/2)\vf}
\end{pmatrix} \ ,
\eeq
while $\(2-\e_{n,+m_j}-\e_{n,-m_j}\) $ is the sum of bound energies of two discrete levels of the same system (\ref{f41b}) for $\pm m_j$, corresponding to the same radial quantum number $n$.

Such approach to evaluation of $ \E_{VP} $ turns out to be quite effective, since $ \d_{tot,|m_j|}(k) $ behaves both in IR and UV-limits in the $k$-variable much better, than each of the scattering phase shifts, considered separately (see below). Moreover, $ \d_{tot,|m_j|}(k) $ will be by construction an even function of the external field. In turn, in the total bound states energy  the condensation point $ \e_{n,m_j} \to 1 $ turns out to be regular for each  $m_j$. Therefore the representation of $ \E_{VP} $ in the form (\ref{f41}) permits to avoid an intermediate regularization of the Coulomb asymptotics of the external potential for $ r \to \inf $, what significantly simplifies all the subsequent calculations.

As a result,   in 2+1 D, as well as in 1+1 D, for the external potentials of the type (\ref{1.00}) each term of the sum over $ m_j $ in the expression for $ \E_{VP} $ turns out to be finite without any special UV-renormalization. This statement follows directly from  (\ref{f41}), since, as it will be shown below via explicit calculation,  $ \d_{tot,|m_j|}(k) $ is regular for $ k \to 0 $ and behaves like  $ \mathrm{O}(1/k^3) $ for $ k \to \inf $, whence the phase integral in (\ref{f41}) turns out to be always convergent, while the total bound states energy is also finite, because  $ 1-\e_{n, \pm m_j} $ behave for $ n \to \inf $ as $ \mathrm{O}(1/n^2) $.

At the same time, there is a principal difference between two-dimensional and one-dimensional problems with the same model potential, the essence of which is that  both $ \r_{VP} $ and $ \E_{VP} $ are represented now as infinite partial expansions in $m_j$.  So there appears a natural question of convergence of these series. The convergence of the partial expansion for $ \r_{VP} $ has been demonstrated in the part I. For the same answer concerning the expansion for $ \E_{VP} $  let us analyze the behavior of  separate terms in the series  (\ref{f41a})  for $|m_j|\to\inf$.
The main component  $ \sim(Z\a)^2$ of the total scattering phase $\d_{tot,|m_j|}$ for large $|m_j| \gg Z\a$ can be found via quasiclassical (WKB) approximation:
\beq
\label{int_WKB}
\bal
&\d_{WKB,|m_j|}(k) \\
&= 2\int \!   dr \ \(\sqrt{\(\e+V(r)\)^2-1-{m_j^2 \over r^2}} +  \sqrt{\(\e-V(r)\)^2-1-{m_j^2 \over r^2}}-2 \sqrt{k^2-{m_j^2 \over r^2}}\) \ ,
\eal
\eeq
where $\e=\sqrt{k^2+1}$, while the integration is performed over regions, where the expressions under the square root are non-negative. The analytic calculations lead to the following result
\beq
\begin{aligned}
\displaystyle&\d_{WKB,|m_j|}(k)=
\\&\pi\(|m_j|-\sqrt{m_j^2-(Z\a)^2}\)(\tt(k_1-k)+\tt(k_2-k)) \ +
\frac{2\e{Z}\a}{\sqrt{\epsilon^2-1}}\ln\(\sqrt{m_j^2+\frac{(Z\a)^2}{\e^2-1}}\)\\
&\times(\tt(k_1-k)-\tt(k_2-k))+2\Bigg[\frac{\e{Z}\a}{\sqrt{\e^2-1}}\ln\(\sqrt{\e^2-1}R-\e Z \a/\sqrt{\e^2-1}\right.\\
&\left. +
\sqrt{(\e^2-1)R^2-2R\e Z\a+(Z\a)^2-m_j^2}\)+\sqrt{m_j^2-(Z\a)^2}\\
&\times\arcsin\(\frac{(Z\a)^2-m_j^2-\e Z \a R}{R\sqrt{(\e^2-1)m_j^2+(Z\a)^2}}\)+ m_j\arctg\(\frac{m_j}{\sqrt{(\e R-Z\a)^2-R^2-m_j^2}}\)\Bigg]\\
&\times\tt(k-k_1)-2\Bigg[\frac{\e Z \a}{\sqrt{\e^2-1}}\ln\(\sqrt{\e^2-1}R+\e Z\a/\sqrt{\e^2-1}\right.\\ \nonumber \eal\eeq
\beq \bal
&\left.+\sqrt{(\e^2-1)R^2+2R\e{Z}\a+(Z\a)^2-m_j^2}\)-\sqrt{m_j^2-(Z\a)^2}\\
&\times\arcsin\(\frac{(Z\a)^2-m_j^2+\e Z \a{R}}{R\sqrt{(\e^2-1)m_j^2+(Z\a)^2}}\)-m_j\arctg\(\frac{m_j}{\sqrt{(\e R+Z\a)^2-R^2-m_j^2}}\)\Bigg]\\
&\times\tt(k-k_2) \ ,
\end{aligned}
\eeq
with $k_1=\sqrt{\(V_0 +\sqrt{1+m_j^2/R^2}\)^2-1}$ and $k_2=\sqrt{\(-V_0+\sqrt{1+m_j^2/R^2}\)^2-1}$ being the quasiclassical turning points.

For comparison in the Fig.1  there are shown the curves of the exact total phase $\d_{tot,|m_j|}(k)$ and of its WKB-approximation $\d_{WKB,|m_j|}(k)$  for $ Z=1000 $ and $ |m_j|=31/2 $.
\begin{figure}[h!]
	\begin{minipage}{0.48\linewidth}
		\center{ \includegraphics[scale=0.7]{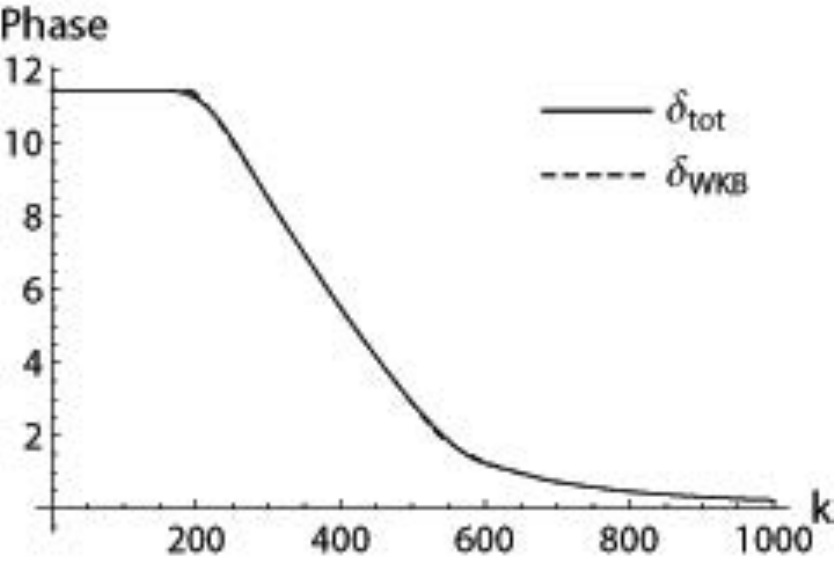}  }
	\end{minipage}
	\hfill
	\begin{minipage}{0.48\linewidth}
		\center{ \includegraphics[scale=0.7]{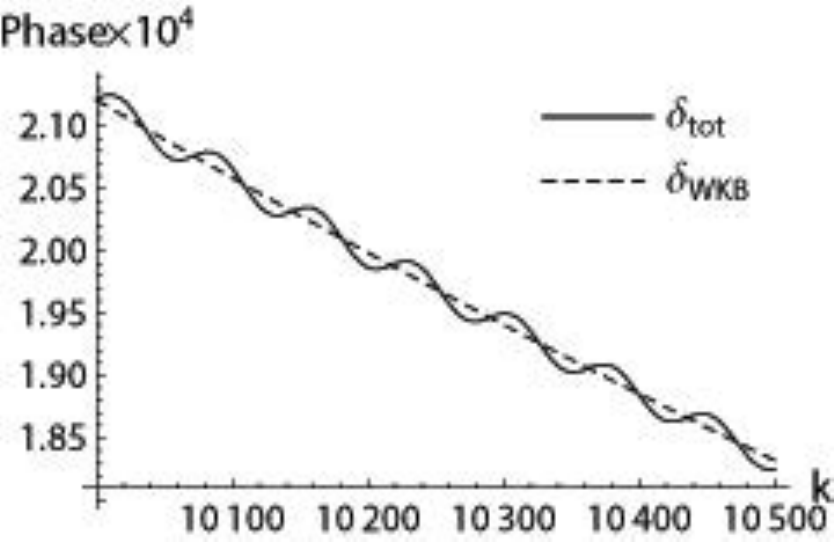}  }
	\end{minipage}
	\begin{center}
		{\small Fig.1. The exact total scattering phase and its WKB-approximation for $Z=1000, \ |m_j|=31/2$.}
	\end{center}
\end{figure}
Let us note that the quasiclassical approximation does not reproduce the oscillations of the exact phase for large  $ k $, which correspond to diffraction on a sphere of the radius  $R$. At the same time, the behavior of $ \d_{tot,|m_j|}(k) $ and $ \d_{WKB,|m_j|}(k) $ for $kR \ll |m_j|$ can be understood by calculating the corresponding total phase for the point-like Coulomb source with the potential $V(r) = -Q/r$ (for $|m_j|>Q$). The analytic solution of the corresponding two-dimensional Dirac equation gives the following answer for each of the  partial phase shifts
\beq\label{pointsource0}
	\delta^{\pm}_{|m_j|}(k)=\frac{\pi} {2}|m_j| \pm \frac{\e Q}{ k}\ln{2 k r} - \frac{\pi \vk} {2} \mp  \1/2 \mathrm{Arg}\[ \G(1+\vk + i \e Q/k)\] + \1/2 \mathrm{Arg} \[ \frac{ |m_j|+i Q/k} { \vk \mp i \e Q/k} \] \ , \eeq
where  $\e=\sqrt{k^2+1}$, $\vk=\sqrt{m_j^2-Q^2}$,  the signs $\pm$ correspond to the phase shifts for the upper and lower continua, while the phases $\delta^{\pm}_{-|m_j|}$ are obtained from $\delta^{\pm}_{|m_j|}$ via simple change of the sign $|m_j|\to -|m_j|$ in the last term in (\ref{pointsource0}).

In this case the total phase  $\d_{tot,|m_j|}(R \to 0)=\sum \d^{\pm}_{\pm |m_j|}(k) $  does not depend on $k$ at all
\beq\label{pointsource}
\d_{tot, |m_j|}(R \to 0)= 2\pi\(|m_j|-\sqrt{m_j^2-Q^2}\) \ ,
\eeq
and coincides exactly both with the exact and the WKB phases for $k R \ll |m_j|$, since under such conditions  there takes place in fact the scattering on the purely Coulomb potential without the central sphere with radius $R$. Note also that for such behavior of the exact phase  the WKB-condition $|m_j| \gg Z\a$ is crucial, otherwise $ \d_{tot,|m_j|}(k) $ for $k \to 0$ will be still finite, but its limiting value in this case can be sufficiently different from (\ref{pointsource}), especially in the case $|m_j|<Z\a$ (see formulae (\ref{f45-16})-(\ref{f45-15}) below). So the smaller $R$, the greater the value of $k$ is needed (more exactly, the correct condition reads $k R \gg |m_j|$) to alter the behavior of the exact and quasiclassical phases from the constant value (\ref{pointsource}) into decrease to zero, smooth in the case of the WKB-approximation and oscillating for the exact phase. Moreover, the result (\ref{pointsource}) shows  that for the  point-like Coulomb source the  method under consideration for calculating the vacuum energy, based on transformation of the contribution from the continua into the phase integral, is not valid, since the total phase in this case becomes a constant, independent of $k$.

To evaluate the limiting behavior of the phase integrals in (\ref{f41a}) for $|m_j|\to\infty$ let us insert into them the integral representation of the scattering  phase in the WKB-approximation (\ref{int_WKB}) by introducing an intermediate UV-cutoff in the energy variable, that provides the possibility of exchange the sequence of integrations

\beq\label{phaseint}
\begin{aligned}
\displaystyle	&I_p(|m_j| \to \inf)=\int\limits_0^\inf \!  \frac{k \, dk }{\sqrt{k^2+1}} \ \d_{tot,|m_j|}(k)\\
&=2\int\limits_0^\inf \! dr \ \lim\limits_{\L \to \infty} \int\limits_1^\L \! d\e \ \Bigg( \sqrt{\(\e+V(r)\)^2-1-m_j^2 /r^2} \ +
 \sqrt{\(\e-V(r)\)^2-1-m_j^2/r^2} \\
 &- 2 \sqrt{\e^2-1-m_j^2/r^2} \Bigg) \ .
\end{aligned}
\eeq
In (\ref{phaseint}) the integration over $d\e$ is performed over the regions from the interval $(1,\L)$, where the expressions under the square root are non-negative. Let us analyze now the position of the turning points, taking into account that the limit $ |m_j| \to \infty$ for fixed  $Z$ is considered. For the latter analysis it turns out to be convenient to use the following subsidiary parameter
\beq\label{r0}
r_0={ m_j^2-(Z\a)^2 \over 2Z\a} \ .
\eeq
For the first integral in (\ref{phaseint}) the turning points are $\e_1^{\pm}(r) = -V(r)\pm\sqrt{1+ m_j^2 /r^2}$. Subject to condition $|m_j| \gg Z\a$ these turning points should always satisfy the relations: $\e_1^+(r) >1,\e_1^-(r)<1$. For the second integral the turning points are $\e_2^{\pm}(r) = V(r)\pm\sqrt{1+ m_j^2/r^2}$. The turning point $\e_2^-(r)$ for any $r$ satisfies the condition $\e_2^-(r)<0$.  Provided $|m_j| > \sqrt{(Z\a)^2+2Z\a R}$ and $r < r_0 $ the turning point $\e_2^+(r)$ satisfies the relation $\e_2^+(r) >1$, while for $|m_j| > \sqrt{(Z\a)^2+2Z\a R}$ and  $r > r_0 $ --- the relation  $\e_2^+ (r)<1$. The turning points of the last integral $\e_0^{\pm}(r) = \pm\sqrt{1+ m_j^2/r^2}$ satisfy always the condition $\e_0^{+}(r)>1,\e_0^-(r)<0$. As a result, the quasiclassical estimate for the phase integral takes the following form
\beq\label{phaseint1}
\begin{aligned}
& I_p(|m_j| \to \inf)= 2\int\limits_0^\inf \! dr \ \lim\limits_{\L \to \infty}\Bigg[ \int\limits_{\e_1^+}^\L \! d\e \ \sqrt{\(\e+V(r)\)^2-1-{m_j^2 \over r^2}} \\
&+ \tt\(r_0-r\) \int\limits_{\e_2^+}^\L \! d\e \ \sqrt{\(\e-V(r)\)^2-1-{m_j^2 \over r^2}}+ \tt\(r - r_0\) \int\limits_{1}^\L  \! d\e \  \sqrt{\(\e-V(r)\)^2-1-{m_j^2 \over r^2}} \\
&- 2\int\limits_{\e_0^+}^{\L} \! d\e \  \sqrt{\e^2-1-{m_j^2 \over r^2}}\Bigg]
\\ &= \int\limits_0^\inf \! dr \ \lim\limits_{\L \to \infty}
\Bigg[(\L+V(r))\sqrt{(\L+V(r))^2-1-{m_j^2 \over r^2}}-\(1+{m_j^2 \over r^2}\)\\
&\times \ln{\Bigg(\L+V(r)+\sqrt{(\L+V(r))^2-1-{m_j^2 \over r^2}} \Bigg)} + (\L-V(r))\sqrt{(\L-V(r))^2-1-{m_j^2 \over r^2}}\\
&-\(1+{m_j^2 \over r^2}\)\ln{\Bigg(\L-V(r)+\sqrt{(\L-V(r))^2-1-{m_j^2 \over r^2}}\Bigg)}- \L\sqrt{\L^2-1-{m_j^2 \over r^2}}\\
&+\(1+{m_j^2 \over r^2}\)\ln{\Bigg(\L+\sqrt{\L^2-1-{m_j^2 \over r^2}}\Bigg)}\Bigg]\\
& + 2\int\limits_{r_0}^{\infty} \! dr \ \(1+{m_j^2 \over r^2}\) \[\ln\(1-V(r)+\sqrt{\(1-V(r)\)^2-1-{m_j^2 \over r^2}}\)\right.\\
&\left. -{1-V(r) \over 1+m_j^2/ r^2}\sqrt{\(1-V(r)\)^2-1-{m_j^2 \over r^2}}-\ln\sqrt{1+{m_j^2 \over r^2}}\] \ .
\eal
\eeq
Proceeding further by calculating the limit $\L \to \inf$ in the first integral and the second integral  in (\ref{phaseint1}), one obtains finally
\beq\label{phaseint2}
\begin{aligned}
& I_p(|m_j| \to \inf)=2\int\limits_0^{\infty} \! dr \ V^2(r)-2\pi\(|m_j|-\sqrt{m_j^2-(Z\a)^2}\)\\
&= 2\int\limits_0^{\infty} \! dr \ V^2(r)-{\pi (Z\a)^2 \over |m_j|} - {\pi (Z\a)^4 \over 4 |m_j|^3}+\mathrm{O}\(1 \over |m_j|^5\) \  .
\end{aligned}\eeq
Taking into account that the next-to-leading order of the WKB-approximation for the total phase leads to corrections, whose  leading terms of expansion in  $1/|m_j|$ for $|m_j| \to \inf$ should be proportional to $(Z\a)^4/|m_j|^3$, there follows from (\ref{phaseint2}) that for large $|m_j| \gg Z\a$ the phase integral should behave as follows
\beq
\int\limits_0^\inf \!   \  \frac{k \, dk }{\sqrt{k^2+1}} \ \d_{tot,|m_j|}(k) = 2\int\limits_0^{\infty} \! dr \ V^2(r)-{\pi (Z\a)^2 \over |m_j|} + \mathrm{O}\(1 \over |m_j|^3\) \ , \quad |m_j|\to \inf \ .
\eeq

At the same time, the discrete levels with the same conditions on  $|m_j|$ (that means $|m_j|\to \inf$ or at least $|m_j| \gg Z\a$) correspond with a high precision to the solutions of two-dimensional Schroedinger equation with the same external potential (\ref{1.00}), including the relativistic corrections \cite{yang1991}, as well as the one, caused by the non-vanishing  size of the Coulomb source
\beq
\label{sum_d}
\bal
1-\e_{n,\pm m_j}&={(Z\a)^2 \over 2(n+|m_j|)^2} \(1+{(Z\a)^2 \over (n+|m_j|)^2}\({n+|m_j| \over |m_j|} - {3 \over 4}\)\right.\\
&\left.-\(2Z\alpha R \over n+|m_j|\)^{2|m_j|}{(n+2|m_j|-1)! \over n!}{2|m_j| (2|m_j|+1) \over ((2|m_j|)!)^2}\) \ , \quad |m_j| \to \inf \ .
\eal
\eeq
In (\ref{sum_d}) for large $|m_j|$ the correction from  the source size turns out to be negligibly small (in fact, exponentially small) compared to the fine-structure one  and so can be omitted. In the next step,  the sum over all discrete levels (\ref{sum_d}) for the fixed  $|m_j|$ is  evaluated by taking into account the following circumstance. Namely, by definition $ (2-\e_{n,+m_j}-\e_{n,-m_j}) $ is the sum of bound energies of two discrete levels of the system (\ref{f41b}), corresponding to  $\pm m_j$ and the same radial number $n$. However, it can be easily verified that in the system (\ref{f41b}) for each $|m_j|$ the lowest level with $n=0$ exists only for $m_j>0$, while for $m_j<0$ the discrete set starts from $n=1$. At the same time, for the mirror-symmetrical system with opposite signature of two-dimensional Dirac matrices or, equivalently, for another subsystem in the four-dimensional representation, which is related to (\ref{f41b}) via the change of the sign $m_j\to -m_j$, the lowest level with $n=0$ exists for $m_j<0$. This effect is in a very close connection with 3+1 D, when in the relativistic hydrogen ion the degeneracy of states with $j=n-1/2$, where $n$ is now the principal quantum number, is twice less than of the others \cite{itzykson1980}. In particular, the lowest $1s_{1/2}$ state with $n=1\, , \ j=1/2$ is degenerated only doubly by the spin projection $\pm 1/2$, while for $n=2$ there exist already two  degenerate states $2s_{1/2}$ and $2p_{1/2}$ of opposite parity, and so the total degeneracy of the energy eigenstate with $n=2\, , \ j=1/2$ is equal to 4. The next level $2p_{3/2}$ is again twice less degenerate than the subsequent $3p_{3/2}$ and $3d_{3/2}$, and so on.

So in fact in (\ref{f41a}) by omitting the common factor $1/2$ the contribution  of discrete levels to $\E_{VP}$ from each partial channel with the fixed $|m_j|$  should be written more carefully, namely
\beq\label{sum_dm}
\bal
& 2 \sum\limits_{-1 \leq \e_{n,\pm m_j}<1} \(2-\e_{n,+m_j}-\e_{n,-m_j}\)
 \\
& = \(2-\e_{0,+m_j}-\e_{0,-m_j}\)+ 2 \sum\limits_{n \geq 1} \(2-\e_{n,+m_j}-\e_{n,-m_j}\) \ ,
\eal
\eeq
where  the common degeneracy factor 2, which stands  in (\ref{f41a}) in front of the whole partial series over $m_j$,  is in one-to-one-correspondence  with the factor 2 in front of the sums over $n$ in (\ref{sum_dm}), while in the contribution from the lowest level such a factor is absent. This is because for $m_j>0$ the lowest level with $n=0$ exists in the system (\ref{f41b}) only, whereas for $m_j<0$, conversely,  in its counterpart, related via $m_j \to -m_j$. Let us also mention that in fact the levels $\e_{0,\pm m_j}$  as the lowest levels of the system (\ref{f41b}) and of its counterpart coincide
\beq\label{e_0}
\e_{0,\pm m_j}=\e_{0, |m_j|} \ .
\eeq

By means of (\ref{sum_d})-(\ref{e_0})  the sum of bound energies of discrete levels for large $|m_j|$  can be easily calculated analytically
\beq\label{sum_d1}
\bal
&\sum\limits_{-1 \leq \e_{n,|m_j|}<1}\(2-\epsilon_{n, +m_j}-\epsilon_{n, -m_j}\)= \(1-\epsilon_{0, |m_j|}\)\\
&+ (Z\a)^2 \(\psi^{(1)}(|m_j|+1)-{(Z\a)^2 \over 2\, |m_j|}\p^{(2)}(|m_j|+1)-{(Z\a)^2 \over 8}\p^{(3)}(|m_j|+1)\) \ ,
\eal
\eeq
where $\p^{(n)}(z)=\text{PolyGamma}[n,z]$.
The expansion of r.h.s. of (\ref{sum_d1})  in the inverse powers of $|m_j|$ gives
\beq
\sum\limits_{-1 \leq \e_{n,|m_j|}<1}\(2-\epsilon_{n, +m_j}-\epsilon_{n, -m_j}\)={(Z\a)^2 \over |m_j|}+{(Z\a)^2/ 6+(Z\a)^4/ 4 \over |m_j|^3} + \mathrm{O}\({1 \over |m_j|^5}\) \ .
\eeq
So in the expression  (\ref{f41a}) for $m_j\to\infty$ the members of the partial series in $m_j$  for the vacuum energy tend to the following finite limit $\sim (Z\a)^2$
\beq
\label{term_m}
\E_{VP,|m_j|}={1 \over \pi} \int\limits_0^{\infty} \! dr \  V^2(r)+\mathrm{O}\({1 \over |m_j|^3}\) \ , \quad |m_j|\to \inf \ ,
\eeq
since the terms $\sim (Z\a)^2/|m_j|$ in the contributions from the phase integral and discrete levels cancel each other.

It should be underlined that this result is in complete agreement with the behavior of the partial series for the renormalized vacuum density $\r_{VP}(r)$ (see part I, eqs. (36,40,44)), where the partial terms $\r^{(3+)}_{VP,|m_j|}(r)$ decrease with growing $|m_j|$ as $\mathrm{O}\(1/|m_j|^3\)$ uniformly in $r$. In turn, this circumstance  can be easily understood as a direct consequence of the general Schwinger relation between $\d \E_{VP}$ and $\d A_0^{ext}(r)$ via $\r_{VP}(r)$ (see (\ref{f46}) below).

It follows from (\ref{term_m}) that the partial series in $m_j$ for $\E_{VP}$ (\ref{f41a}) diverges linearly, whence it follows the necessity of its regularization and subsequent renormalization. At the same time, each partial term in (\ref{f41a}) in itself is finite without any additional manipulations. It should be specially noted that  the degree of divergence of the partial series (\ref{f41a}) for $ \E_{VP} $ is formally indeed the same (linear), as within PT in 2+1 D without virtual photons for the  unique divergent Feynman graph in the form of the fermionic loop with two external lines. The latter circumstance shows that by calculation of $ \E_{VP} $ via the principally different non-perturbative approach, that does not reveal any connection with PT, we nevertheless meet actually the same  divergence of the theory, as in PT. In fact,  it should be indeed so, since both approaches deal with the same physical phenomenon (vacuum polarization caused by the strong Coulomb field) with the main difference in  the methods of calculation. And so in the present approach the cancelation of divergent terms should follow the same rules as in PT, based on the regularization of the fermionic loop with two external lines, that preserves the physical essence of the whole renormalization procedure and simultaneously provides the mutual agreement between perturbative and non-perturbative approaches to the calculation of $ \E_{VP} $.  This conclusion is in the complete agreement with results obtained in Ref.~\cite{gyul1975}.

The need in the renormalization via fermionic loop follows also from the analysis of the properties of $\r_{VP}$, which shows that without such UV-renormalization the integral vacuum charge will not acquire the expected integer value in units of $(-2 |e|)$  (see part I, Section 4). In fact, the properties of $\r_{VP}$ play here the role of a controller, that provides the implementation of the required physical conditions for a correct description of the vacuum polarization effects beyond the scope of PT, which cannot be tracked via evaluation of $\E_{VP}$ by means of the initial relations (\ref{f40}),(\ref{f41a}). Moreover, for $Z \to 0$ the vacuum energy should coincide with  $\E^{(1)}_{VP}$, obtained within PT according to (\ref{2.1})-(\ref{2.6}).  Recall, that due to the axial symmetry of the external field both the perturbative density and vacuum energy correspond to the partial channel with $ |m_j|=1/2 $ (see also part I, Appendix A).  It is easy to verify, however, that in the general case the non-renormalized  $ \E_{VP, 1/2}$ does not reproduce the perturbative answer for $Z \to 0$. In particular, for the external field (\ref{1.00}) it is easy to verify by  direct calculation that the analytic answers   for $ \r_{VP}^{(1)}(r) $ (\ref{2.5c}) and for $ \r_{VP,1/2}(r)$,  found from the first Born approximation for the Green function of the DC spectral problem (see part I, formulae (34), (37-39)), hence for $ \E_{VP}^{(1)} $ and $ \E_{VP, 1/2}$,  turn out to be substantially different for $Z \to 0$.

 Thus, in the complete analogy with the renormalization of the vacuum density, considered in  part I  (see I, formulae (36),(40)), we should pass to the renormalized vacuum energy by means of the relation
\beq
\label{f42}
\E^{ren}_{VP}(Z) = 2\sum\limits_{m_j=1/2,3/2,..} \E^{ren}_{VP,|m_j|}(Z) \ , \quad \E^{ren}_{VP,|m_j|}(Z)=\E_{VP,|m_j|}(Z)+ \eta_{|m_j|}(R) Z^2 \ ,
\eeq
where
\beq
\label{f43}
\eta_{|m_j|}(R) = \lim\limits_{Z_0 \to 0}  \[{\E_{VP}^{(1)}(Z_0)\d_{|m_j|,1/2}-\E_{VP,|m_j|}(Z_0) \over Z_0^2}\]_{R=R(Z)}
\ . \eeq
The key-point of (\ref{f42}) is that now from  the  initial expressions for non-renormalized partial terms  $ \E_{VP,|m_j|}(Z) $ in (\ref{f41a}) the quadratic in  $Z$ components are separated and replaced further by the renormalized  $\E^{(1)}_{VP}\d_{|m_j|,1/2}$, found within PT. This procedure is in complete agreement with the renormalization of $\r_{VP}$ with the only difference, that in the latter case the same procedure is applied to the linear in $Z$ components. Moreover, such a renormalization provides simultaneously the convergence of the partial series for  $\E_{VP}^{ren}$, since the divergent terms in the sum (\ref{f41a}), according to (\ref{term_m}), are proportional to $(Z\a)^2$. So the renormalization via fermionic loop turns out to be the universal method, which removes the divergence of the theory both in purely perturbative and essentially non-perturbative approaches to the vacuum polarization.

It should be underlined specially that actually the renormalization coefficients  $ \eta_{|m_j|} $ are determined by the shape of the external field, and so in the present case turn out to be the functions of  the radius $R(Z)$ of the central sphere in the potential (\ref{1.00}), thence of the current charge  $Z$ of the external source.  It is easy to see that they are represented as a double integral of $ A_0^{ext}(r)/Z $, what for one finds firstly  $ \E_{VP}^{(1)}(Z_0) $  by means of (\ref{2.1}), and further  $ \E_{VP,|m_j|}(Z_0) $ by means of the first Born approximation for $ \r_{VP,|m_j|}(r)$ (see part I, formulae (34), (37-39)), with the external potential (\ref{1.00}) with the charge $Z_0$, but with the radius of the central sphere $R=R(Z)$. As a result,
\beq
\label{f44}
\bal
&\E_{VP}^{(1)}(Z_0)\d_{|m_j|,1/2}-\E_{VP,|m_j|}(Z_0) \\
&=  \1/2 \int \! d^2r \ \( Z_0 A_0^{ext}(r)/Z\) \[ \(\r_{VP}^{(1)}\)_{PT}(r)\d_{|m_j|,1/2}-\(\r_{VP,|m_j|}^{(1)}\)_B(r)\] \ ,
\eal
\eeq
where  $ \(\r_{VP}^{(1)}\)_{PT}(r) $ is the renormalized perturbative vacuum density from the first-order PT (\ref{2.2})-(\ref{2.5c}), while $ \(\r_{VP,|m_j|}^{(1)}\)_{B}(r) =(|e|/\pi)\, \int \! dy \ \mathrm{Re}\,\tr G^{(1)}_{m_j}(r, iy)= - (|e|/\pi)\, \int\! dy\  \tr \(G^{(0)}_{m_j} V G^{(0)}_{m_j}\)$ is the vacuum density, found through the first Born approximation (see part I, formulae (34), (37-39)), and both densities in turn are expressed via the integrals of $Z_0 A_0^{ext}(r)/Z $.

So in its final version the renormalization of $\E_{VP}$ reduces to replacing the terms of the partial series in $m_j$ in the expression (\ref{f41a}) by
\beq\label{renenergy}
\bal
&\E^{ren}_{VP,|m_j|}(Z) \\
&= {1 \over 2\pi} \int\limits_0^\inf \!   \  \frac{k \, dk }{\sqrt{k^2+1}} \ \d_{tot,|m_j|}(k) + {1 \over 2} \sum\limits_{-1 \leq \e_{n, \pm m_j}<1} \(2-\epsilon_{n, +m_j}-\epsilon_{n, -m_j}\) \\
& + \eta_{|m_j|}(R(Z)) Z^2 \ .
\eal
\eeq
 It would be worth noticing that actually each partial channel in (\ref{renenergy}) reproduces by its structure almost exactly the renormalized  $\E_{VP}$ in the one-dimensional case \cite{davydov2017,sveshnikov2017,voronina2017}. The whole difference is that $\E_{VP}^{ren}$  in the latter case  contains always  the difference $\(\E_{VP}^{(1)}(Z_0)-\E_{VP}(Z_0)\)/Z_0^2$ in the renormalization coefficient $\eta$, whereas in the present case such a difference resides  only in the partial term with $|m_j|=1/2$. However, in the one-dimensional case $\eta(R)$ is  a nontrivial  sign-alternating function of the radius $R$ \cite{davydov2017,voronina2017}, but in  2+1 D all the  $ \eta_{|m_j|} $, including $ \eta_{1/2} $, turn out to be always strictly negative (see Fig. 6a,b).

\section{Renormalized Vacuum Energy for $Z>Z_{cr,1}$: Explicit Evaluation for the External Field (\ref{1.00})}

Now let us turn to the explicit evaluation of $\E^{ren}_{VP}(Z)$ by means of (\ref{f41a}) and (\ref{renenergy}) for the potential (\ref{1.00}).
For $r\leq R$ the solutions of the system  (\ref{f391}) up to a common normalization factor take the form
\beq
\begin{aligned}
\label{f450}
&\psi^{int}_{1, m_j}(r,\e )=(-i)^{(m_j-1/2)\theta(1-|\epsilon+V_0|)}\, \sqrt{|\epsilon+V_0+1|}\, J_{m_j-1/2}(\z r),\\
&\psi^{int}_{2, m_j}(r, \e )=(-i)^{(m_j+1/2)\theta(1-|\epsilon+V_0|)}(-1)^{\theta(\epsilon +V_0 - 1)}\, \sqrt{|\epsilon+V_0-1|}\, J_{m_j+1/2}(\z r) \ ,
\end{aligned}
\eeq
where $J_{\nu}(z)$ are the Bessel functions,
\beq
V_0=Z \a/R,\quad \z=\sqrt{(\e+V_0)^2-1} \ ,
\eeq
while the phase factors $(-i)^{(m_j \mp 1/2)\theta(1-|\epsilon+V_0|)}$ are inserted to provide the purely real solutions (\ref{f450}) by transition through  the region of the hyperbolic regime $|\epsilon+V_0|<1$, where the Bessel functions are replaced by the  corresponding Infeld ones.

The solutions of the system (\ref{f391}) for  $ r>R $ should be represented now in terms of the Kummer and Tricomi functions $\F(b,c,z)$ and $\P(b,c,z)$ \cite{bateman1953}. In the upper and lower continua for $|m_j| > Q$ these solutions take the following form
\beq
\label{f45}
\begin{aligned}
	\p_{1,\, m_j}^{ext}(r,\e)&= \sqrt{|\e + 1|} \,r^{\vk-1/2}\,\( \mathrm{Re}\[\mathrm{e}^{i\f_+}\mathrm{e}^{i k r} \F_r\]+B_{m_j}(\e)\, \mathrm{Re}\[i \mathrm{e}^{-i\pi\vk}\mathrm{e}^{i\f_-}\mathrm{e}^{i k r} \tilde \F_{r}\] \),
	\\
	\p_{2,\, m_j}^{ext}(r,\e)&= -\sign(\e)\sqrt{|\e - 1|}\, r^{\vk-1/2}\\
	&\times  \( \mathrm{Im}\[\mathrm{e}^{i\f_+}\mathrm{e}^{i k r} \F_r\]+B_{m_j}(\e)\, \mathrm{Im}\[i \mathrm{e}^{-i\pi\vk}\mathrm{e}^{i\f_-}\mathrm{e}^{i k r} \tilde \F_{r}\] \),
\end{aligned}
\eeq
where $\e = \pm\sqrt{k^2 + 1}$ \ ,
\beq
\label{f45-1}
\begin{aligned}
& \vk=\sqrt{m_j^2-Q^2} \ , \qquad b=\vk - i\e Q / k \ , \qquad c=1+2\vk \ , \\
& \f_+=\1/2 \mathrm{Arg}\[{m_j+iQ/k\over b}\], \qquad \f_-=\1/2 \mathrm{Arg}\[{b \over m_j-iQ/k}\] \ , \\
&\F_r=\F\(b,c,-2ikr\) \ , \qquad \tilde \F_r=(-2ikr)^{1-c}\F\(1+b-c,2-c,-2ikr\) \ , \\
\end{aligned}
\eeq
while the coefficients $B_{m_j}(\e)$ are determined by matching the internal and external solutions at the point $r=R$
\begin{equation}
	\label{f45-2}
	B_{m_j}(\e)=-{C_{1,\, m_j}(\e)\, \mathrm{Im}\[\mathrm{e}^{i\f_+}\mathrm{e}^{i k R} \F_{R}\] - C_{2,\, m_j}(\e)\, \mathrm{Re}\[\mathrm{e}^{i\f_+}\mathrm{e}^{i k R} \F_{R}\] \over C_{1,\, m_j}(\e)\, \mathrm{Im}\[i \mathrm{e}^{-i\pi\vk}\mathrm{e}^{i\f_-}\mathrm{e}^{i k R} \tilde \F_{R}\] - C_{2,\, m_j}(\e)\, \mathrm{Re}\[i \mathrm{e}^{-i\pi\vk}\mathrm{e}^{i\f_-}\mathrm{e}^{i k R} \tilde \F_{R}\]} \ ,
\end{equation}
where
\beq
\label{f45-3}
C_{1,\,m_j}(\e)=-\sign(\e)\sqrt{|\e - 1|}\ \psi^{int}_{1, m_j}(R,\e ) \ , \quad
C_{2,\,m_j}(\e)=\sqrt{|\e + 1|}\ \psi^{int}_{2, m_j}(R,\e ) \ .
\eeq
For $|m_j| < Q$ the corresponding solutions for  $ r>R $ are written as
\beq
\label{f45-5}
\begin{aligned}
	\p_{1,\, m_j}^{ext}(r,\e)&=\sqrt{|\e + 1|}  \, \mathrm{Re}\[\mathrm{e}^{i\l_{m_j}(\e)} \mathrm{e}^{ikr}(2 k r)^{i|\vk|-1/2}\((m_j+i Q/k)\F_r+b \F_r(b+)\)\] \ ,
	\\
	\p_{2,\, m_j}^{ext}(r,\e)&=-\sign(\e)\sqrt{|\e - 1|} \, \mathrm{Re}\[i \,\mathrm{e}^{i\l_{m_j}(\e)} \mathrm{e}^{ikr}(2 k r)^{i|\vk|-1/2}\right.\\
	&\left.\times\(-(m_j+i Q/k)\F_r+b \F_r(b+)\)\] \ ,
\end{aligned}
\eeq
where now
\beq
\label{f45-6}
\begin{aligned}
	&|\vk|=\sqrt{Q^2-m_j^2} \ , \qquad b=i\(|\vk| -\e Q / k \),\qquad c=1+2i|\vk| \ , \\
	& \F_r=\F\(b,c,-2ikr\) \ , \qquad \F_r(b+)=\F\(b+1,c,-2ikr\) \ .
\end{aligned}
\eeq
The coefficients $\lambda_{m_j}(\e)$ are found by matching the solutions (\ref{f45-5}) with corresponding ones from the region $r<R$
\begin{equation}
	\label{f45-7}
	\bal
\lambda_{m_j}(\e)&= -\mathrm{Arg}\left[i \mathrm{e}^{i k R}(2 k R)^{i |\varkappa|}\left( (C_{2,\, m_j}+i C_{1,\, m_j})(m_j+iQ/k)\F_{R}\right.\right.\\
&\left.\left.+(C_{2,\, m_j}-i C_{1,\, m_j})b\F_{R}(b+) \right)\right] \ .
	\eal
\end{equation}

The discrete spectrum is determined from conditions of vanishing solutions at the spatial infinity combined with their matching at the point $r=R$. Since there holds now $|\e|<1$, the solutions of the system (\ref{f391})  for $r>R$ are written via $\F(b,c,z)$ and  $\P(b,c,z)$ in a different fashion. Namely, if
\beq
\g=\sqrt{1-\e^2} \ , \quad z=2 \g r \ ,
\eeq
then for  $|m_j|>Q$ the most convenient form of representing the solutions is realized via the Tricomi function $\P(b,c,z)$
\beq\bal
&\psi^{ext}_{1, m_j}(r,\e)= \sqrt{1+\e} \,\mathrm{e}^{-\g r} r^{-1/2+\vk} \[ \Psi + \( Q/\g - m_j \) \Psi(b+)\] \ , \\
&\psi^{ext}_{2, m_j}(r,\e)= \sqrt{1-\e} \,\mathrm{e}^{-\g r} r^{-1/2+\vk} \[ -\Psi + \( Q/\g - m_j \) \Psi(b+)\] \ ,
\eal\eeq
where $b=\vk-\e Q/\g$ , $ c=1+2\vk$, while the equation, determining the discrete levels, takes the form
\begin{multline}
	\label{f45-8}
	\sqrt{(\e+V_0+1)(1-\e)}\,J_{m_j-1/2}(\z R)\,\[-\P+(Q/\g-m_j)\,\P(b+)\]\\
	+\sqrt{(\e+V_0-1)(1+\e)}\,J_{m_j+1/2}(\z R)\,\[\P+(Q/\g-m_j)\,\P(b+)\]=0 \ .
\end{multline}
At the same time, for $|m_j|<Q$ the most correct form of representing the solutions of the system (\ref{f391}) for $r>R$ is achieved via the Kummer function $\F(b,c,z)$
\beq\bal
&\psi^{ext}_{1, m_j}(r,\e)= \sqrt{1+\e} \,\mathrm{e}^{-\g r} \mathrm{Re}\[ \mathrm{e}^{i\l}(2 \g r)^{-1/2+i |\vk|} \( \( Q/ \g + m_j \) \Phi +  b\, \Phi(b+)\)\] \  , \\
&\psi^{ext}_{2, m_j}(r,\e)=\sqrt{1-\e} \,\mathrm{e}^{-\g r} \mathrm{Re}\[\mathrm{e}^{i\l} (2 \g r)^{-1/2+i |\vk|} \( -\( Q/ \g + m_j \)\Phi +  b\, \Phi(b+)\)\] \ ,
\eal\eeq
where $b=i |\vk|-\e Q/ \g \ ,  \quad c=1+2 i |\vk|$. Here the phase  $\l$ is determined by matching the internal and external solutions, while the equation for the discrete levels follows  from the condition of vanishing solutions at the spatial infinity $r\to \infty$ and is represented in the following form
\label{f45-9}
\begin{multline}
	\mathrm{Im}\Big[(2\g R)^{i|\vk|}\, \Gamma(c^*)\Gamma(b)\\
	\times\Big(\sqrt{(\e+V_0+1)(1-\e)}\,J_{m_j-1/2}(\z R)\,\(-(Q/\g+m_j)\F+b\, \F(b+)\)\\
	+\sqrt{(\e+V_0-1)(1+\e)}\,J_{m_j+1/2}(\z R)\,\((Q/\g+m_j)\F+b\, \F(b+)\)\Big)\Big]=0 \ .
\end{multline}

The total phase $\delta_{tot,|m_j|}$, including the contributions from both continua $(\pm)$ and $\pm |m_j|$, as in (\ref{pointsource}), by definition is given by the sum
\beq
\label{f45-10}
\delta_{tot,|m_j|}(k)=\(\delta^+_{|m_j|}+\delta^+_{-|m_j|}+\delta^-_{|m_j|}+\delta^-_{-|m_j|}\)(k) \ ,
\eeq
in which the separate phase shifts $\delta^\pm_{\pm|m_j|}(k)$ are found from the asymptotics of solutions (\ref{f45}) or (\ref{f45-5}) for $r \to \inf$ and contain the Coulomb logarithms $\pm Q\, (|\e|/k)\, \ln (2 k r)$, which cancel mutually in the total phase  (\ref{f45-10}) and henceforth will be omitted in the expressions for separate phases (\ref{f45-11}),(\ref{f45-12}).

As  a result, for $|m_j|>Q$ the phase shifts without the Coulomb logarithms take the form (up to additional  $\pi n$)
\beq
\label{f45-11}
\begin{aligned}
	\delta_{m_j}=\mathrm{Arg}\[\mathrm{e}^{(\pi i/2)|m_j|} \( {\mathrm{e}^{i\f_+}\mathrm{e}^{-i \pi\vk/2} \over \Gamma(1+b^*)} + i B_{m_j}(\e) {\Gamma(2-c)\over \Gamma(c)} {\mathrm{e}^{i\f_-}\mathrm{e}^{i \pi\vk/2} \over  \Gamma(1-b)}\)\] \ ,
\end{aligned}
\eeq
while for $|m_j|<Q$
\beq
\label{f45-12}
\begin{aligned}
	\delta_{m_j}=\mathrm{Arg}\[\mathrm{e}^{(\pi i/2)|m_j|} \( {(m_j+i Q/k)\Gamma(c) \over \Gamma(c-b)}\,\mathrm{e}^{i\lambda_{m_j}(\e)}\mathrm{e}^{\pi|\vk|} +  {\Gamma(c^*)\over \Gamma(b^*)} \,\mathrm{e}^{-i\lambda_{m_j}(\e)}\)\] \ .\\
\end{aligned}
\eeq
 It should be specially noted that the additional $\pi n$ in the phase shifts are in principle unavoidable, since this arbitrariness originates from the possibility to alter the  common factor in the wavefunctions. So here one needs to apply a special procedure, which provides to distinguish between artificial jumps in the phases by  $\pi$ of purely mathematical origin that is inherent in the inverse $\tan$-function, and the physical ones, which are caused by resonances and for extremely narrow low-energy resonances look just like the same jumps by  $\pi$.  Removing the first ones, coming from the inverse $\tan$-function, we provide   the continuity of the phases, while the latter contain an important physical information, hence, no matter how narrow they might be, they must necessarily be preserved in the phase function, and so this procedure should be performed with a very high level of accuracy.

It should be remarked also that after the extraction of Coulomb logarithms the separate phase shifts contain still the singularities for both $k \to 0$ and $k \to \inf$.  Namely, for $k\to\infty$ in the asymptotics of separate phases the singular terms $\mp Q|\e|\ln(2 k R)/k$ are present, but they disappear in the total phase. As  a result, the asymptotic  behavior of $\d_{tot,|m_j|}$ for $k \to \inf$ takes the form
\beq
\label{f45-18}
\bal
&\delta_{tot,|m_j|}(k\to \infty)\\
&={Q\over R^3 k^3}\({4 Q\over 3}(m_j^2-3R^2)-|m_j|\cos(2Q+\pi |m_j|)\sin(2 k R)\)+O(1/k^4) \ .
\eal
\eeq
Let us note that the derivation of the asymptotics (\ref{f45-18}) with account for next-to-leading orders of expansion in $1/k$ for a reasonable time is possible only be means of symbolic computer algebra tools.

 The IR asymptotics of separate phases contain also the singularities of the form $\pm Q/k(1-\ln(Q/k))$. However, these singularities again cancel each other in  $\delta_{tot,|m_j|}$, and so the total phase for  $k \to 0$ possesses a finite limit, which for large $|m_j| \gg Q$ coincides with the WKB-approximation and reproduces the answer for a  point-like source, but in general case turns out to be quite different, especially for $|m_j| <Q$. Namely, for $|m_j|>Q$ the exact limiting value of the total phase for $k \to 0$ equals to
  \beq
\label{f45-16}
\delta_{tot,|m_j|}(k\to 0)=\mathrm{Arg}\[-\mathrm{e}^{-2i\pi\varkappa}v_{1,+} v_{1,-} v_{2,+} v_{2,-}\] \ ,
\eeq
where the following definitions are used
\beq
\label{f45-17}
\bal
v_{1,\pm}&=J_{|m_j|\mp1/2}(R\sqrt{V_0(V_0+2)})\,\(\mp J_{-2\varkappa}(\sqrt{8 Q R})\pm \mathrm{e}^{2i\pi\varkappa}J_{2\varkappa}(\sqrt{8 Q R})\)\\
&\times\sqrt{(V_0+2)/ V_0}+J_{|m_j|\pm1/2}(R\sqrt{V_0(V_0+2)})\\
&\times\Big[\(\sqrt{2 Q R}J_{1+2\varkappa}(\sqrt{8 Q R})+(\mp|m_j|-\varkappa)J_{2\varkappa}(\sqrt{8 Q R})\)\mathrm{e}^{2i\pi\varkappa}\\
	&-\(\sqrt{2 Q R}J_{1-2\varkappa}(\sqrt{8 Q R})+(\mp|m_j|+\varkappa)J_{-2\varkappa}(\sqrt{8 Q R})\)\Big]/Q \ ,\\
 v_{2,\pm}&=\mathrm{Im}\[(-i)^{(|m_j|\mp1/2)\theta(2-V_0)}J_{|m_j|\mp1/2}(R\sqrt{V_0(V_0-2)})\right.\\
&\left.\times\(J_{-2\vk}
 (\sqrt{-8 Q R})\mathrm{e}^{-i\pi\vk}-J_{2\vk}(\sqrt{-8 Q R})\mathrm{e}^{i\pi\vk}\)\] \ .
 \eal
\eeq
For the case $|m_j|<Q$ the corresponding limit could be represented as the following one
\beq
\label{f45-13}\begin{aligned}
\delta_{tot,|m_j|}(k\to 0)&=\mathrm{Arg}\[-\(\mathrm{e}^{\pi|\vk|}\mathrm{e}^{i\varphi^+_{|m_j|}}
-\mathrm{e}^{-\pi|\vk|}\mathrm{e}^{-i\varphi^+_{|m_j|}}\)\right.\\
&\left.\times\(\mathrm{e}^{\pi|\vk|}\mathrm{e}^{i\varphi^+_{-|m_j|}}
-\mathrm{e}^{-\pi|\varkappa|}\mathrm{e}^{-i\varphi^+_{-|m_j|}}\)\sin(\varphi^-_{|m_j|})\sin(\varphi^-_{-|m_j|})
\] \ ,
\end{aligned}\eeq
where the additional phases $\varphi^\pm_{\pm|m_j|}$ are determined through
\beq
\label{f45-14}
\begin{aligned}
	\varphi^+_{\pm|m_j|}&=-\mathrm{Arg}\[\pm\sqrt{2QR}J_{1+2 i|\varkappa|}(\sqrt{8QR})J_{|m_j|\pm1/2}(R\sqrt{V_0(V_0+2)})\right.\\
	&\left.+J_{2 i|\varkappa|}(\sqrt{8QR})w^+_{\pm|m_j|}\],\\
	\varphi^-_{\pm|m_j|}&=-\mathrm{Arg}\Big[(-i)^{(|m_j|-1/2)\theta(2-V_0)}\Big(\sqrt{-2QR}J_{1+2 i|\varkappa|}(\sqrt{-8QR})\\
	&\times J_{|m_j|\mp1/2}(R\sqrt{V_0(V_0-2)})\mp J_{2 i|\varkappa|}(\sqrt{-8QR})w^-_{\pm|m_j|}\Big)\Big] \ ,\\
\end{aligned}
\eeq
with the coefficients
\beq
\label{f45-15}
\begin{aligned}
	&w^+_{\pm|m_j|}\\
	&=Q\sqrt{V_0+2\over V_0}J_{|m_j|\mp1/2}(R\sqrt{V_0(V_0+2)})-(|m_j|\pm i|\varkappa|)J_{|m_j|\pm1/2}(R\sqrt{V_0(V_0+2)}) \ ,\\
	&w^-_{\pm|m_j|}\\
	&=Q\sqrt{V_0-2\over V_0}J_{|m_j|\pm1/2}(R\sqrt{V_0(V_0-2)})-(|m_j|\mp i|\varkappa|)J_{|m_j|\mp1/2}(R\sqrt{V_0(V_0-2)}) \ .\\
\end{aligned}
\eeq

The peculiar feature in the behavior of $ \d_{tot,|m_j|}(k) $ is the appearance of (positronic) elastic resonances upon diving of each subsequent discrete level into the lower continuum. In the Fig.2 the curves, demonstrating the  dependence of the total phase with $ |m_j|=1/2 $  on the wavenumber for various values of $Z$, are shown.
\begin{figure}[h!]
	\begin{minipage}{0.48\linewidth}
		\center{ \includegraphics[scale=0.7]{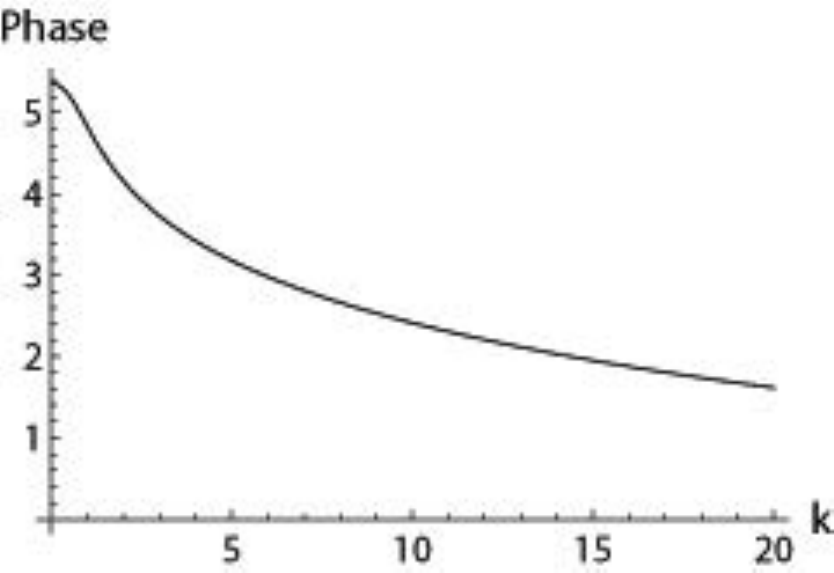} \\ a) }
	\end{minipage}
	\hfill
	\begin{minipage}{0.48\linewidth}
		\center{ \includegraphics[scale=0.7]{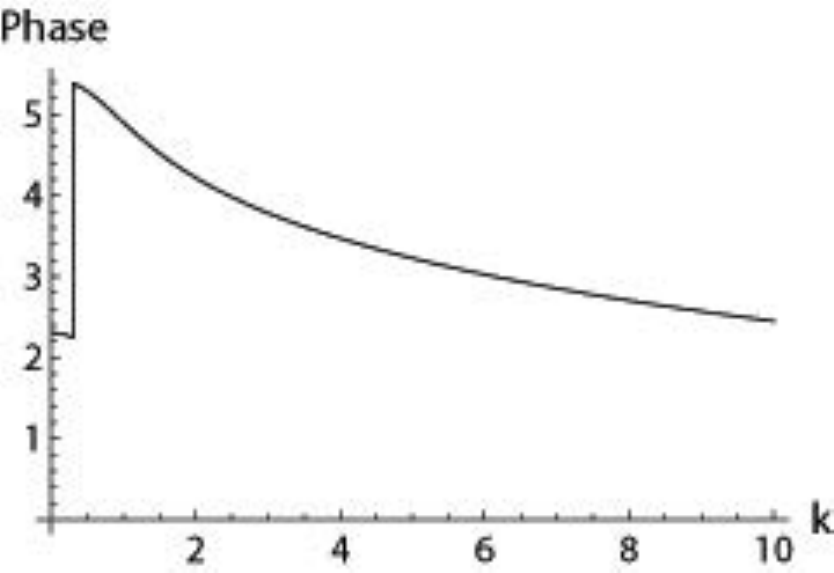} \\ b) }
	\end{minipage}
\end{figure}
\begin{figure}[h!]
	\begin{minipage}{0.48\linewidth}
		\center{\includegraphics[scale=0.7]{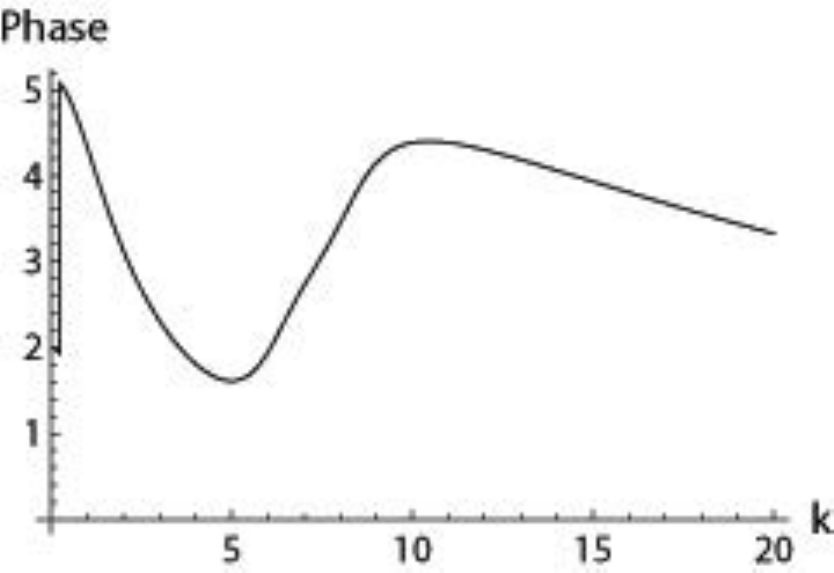} \\ c)}
	\end{minipage}
	\hfill
	\begin{minipage}{0.48\linewidth}
		\center{\includegraphics[scale=0.7]{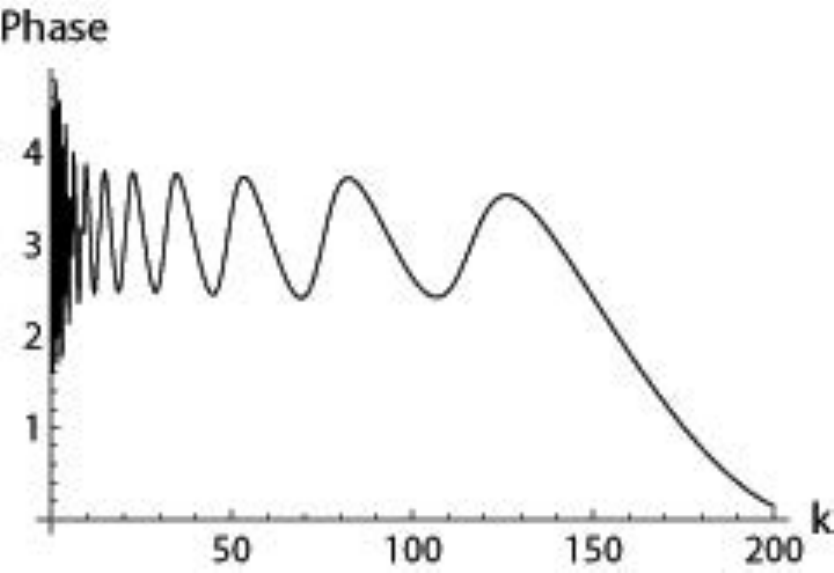} \\ d)}
	\end{minipage}
	\begin{center}
		{\small Fig.2. The dependence of the total phase $ \d_{tot,|m_j|} $ on the wavenumber $k$ for $|m_j|=1/2$ and (a): $ Z=108 $; (b): $ Z=109 $; (c): $Z=193$; (d): $Z=1000$ }.
	\end{center}
\end{figure}

The Fig.2a corresponds to $ Z=108 $, when  no level has reached yet the lower continuum (recall, that $Z_{cr,1}\simeq 108.1$ (see part I)). For $ Z=109 $ (Fig.2b) the first level has already reached the lower continuum, what reflects in the emergence of the first quite narrow low-energy elastic resonance.
With the further growth of $ Z $ up to the next critical value, this jump by $ \pi $ in the phase  is gradually smoothed out and  shifted to the region  of larger  $k$. In the Fig.2c the behavior of the phase for $ Z=193 $ is shown, when already three levels have sunk into the lower continuum, while the last one has reached the lower threshold just now at  $ Z_{cr,3}=192.1$. This level  yields in the phase the first extremely narrow resonance. The two other resonances, originating from two previous levels, lie to the right and reveal substantially less pronounced form. In the Fig.2d the phase for $Z=1000$ is shown, when the number of dived into the lower continuum levels in the channel $ |m_j|=1/2 $ with account for the  degeneracy factor 2 equals to 22 (while the total number of vacuum shells, emerging from the levels, which have sunk into the lower continuum, for all channels  equals to 210).  The details of phase behavior for $Z=1000$ and  $|m_j|=1/2 $ in the regions of small and large $ k $ are shown in Fig.3. In the region of small  $k$ (Fig.3a), the behavior of the phase at the beginning is almost irregular due to jumps, caused by the levels, which have just dived into the lower continuum. At the same time, for large $k$ the phase decreases monotonically with slight  oscillations, that is shown in the Fig.3b. For the other  $ m_j $ the phase behaves quite analogously. So the total phase  $ \d_{tot,|m_j|}(k) $ turns out to be regular everywhere on the whole half-axis $0 \leq k \leq \infty$, while in the region of large $k$ it decreases sufficiently fast to provide the convergence of the phase integral in (\ref{f41}),(\ref{renenergy}), and so the latter could be quite reliable found by means of the standard numerical recipes.

\begin{figure}[h!]
	\begin{minipage}{0.48\linewidth}
		\center{ \includegraphics[scale=0.7]{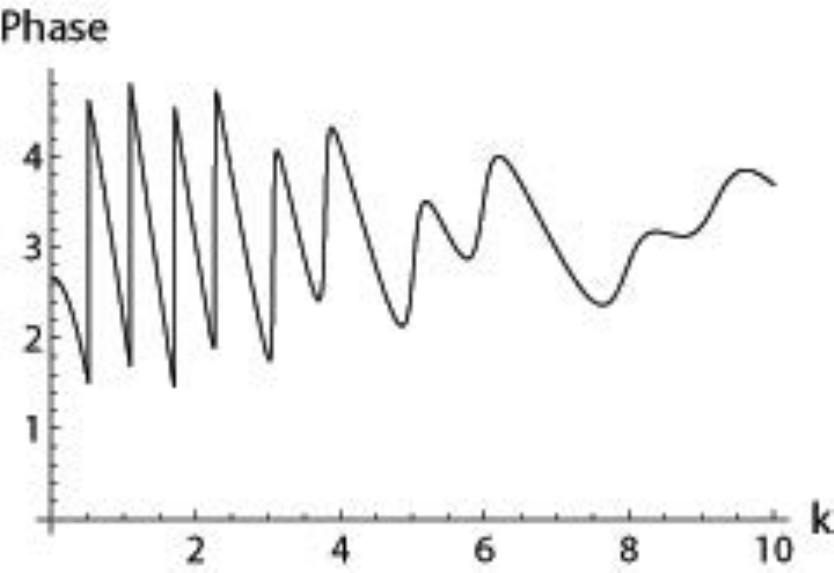} \\ a) }
	\end{minipage}
	\hfill
	\begin{minipage}{0.48\linewidth}
		\center{ \includegraphics[scale=0.7]{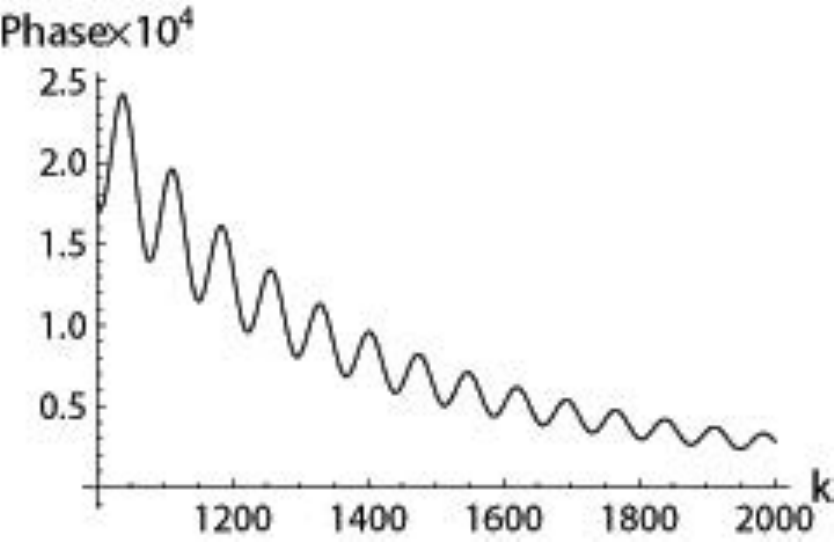} \\ b) }
	\end{minipage}
	\begin{center}
		{\small Fig.3. The dependence of the total phase $ \d_{tot,|m_j|} $ on the wavenumber $k$ for $|m_j|=1/2$ and $Z=1000$}.
	\end{center}
\end{figure}

The typical dependence of the phase integral on $Z$ is presented in the Fig.4. As it follows from  Fig.4, the phase integral increases monotonically as a function of $ Z $ and is always positive. The clearly seen bending, but in fact a jump in the derivative of the curve, takes place at $Z=Z_{cr,1}$, when the first discrete level reaches the lower continuum. At this moment the behavior of the phase integral as a function of $Z$  changes markedly, since there appears a negative jump of the derivative.  The origin of this jump is  that in the total phase, due to the resonance just born, there appears a sharp jump by  $\pi$ (see Fig.2b,c). The subsequent resonances also lead to the jumps in the derivative of the phase integral, but they turn out to be much less pronounced already, since with increasing $Z_{cr}$ there shows up the effect of  ``the catalyst poisoning'' (like sticking of $\m$ to $\alpha$ in the muonic catalysis) --- just below the threshold of the lower continuum for $Z=Z_{cr}+\D Z \ , \ \D Z \ll Z_{cr}$, the  resonance broadening and its rate of diving into the lower continuum behave exponentially slower in complete agreement  with the well-known result  \cite{zeld1971}, according to which the resonance width just under the threshold behaves like $\sim \exp \( - \sqrt{Z_{cr}/\D Z}\)$. This effect leads to that for each subsequent resonance the region of the phase jump by $\pi$ with increasing  $Z$ grows exponentially slower, and derivative of the phase integral changes in the same way. If not for this effect, then each next level reaching the lower continuum would lead to the same negative jump in the derivative as the first one, and the phase integral curve in the overcritical region would have an ever increasing negative curvature with all the ensuing consequences for the rate of decrease of the total vacuum energy $\E_{VP}^{ren}(Z) $.
\begin{figure}[h!]
	\begin{minipage}{0.48\linewidth}
		\center{ \includegraphics[scale=0.7]{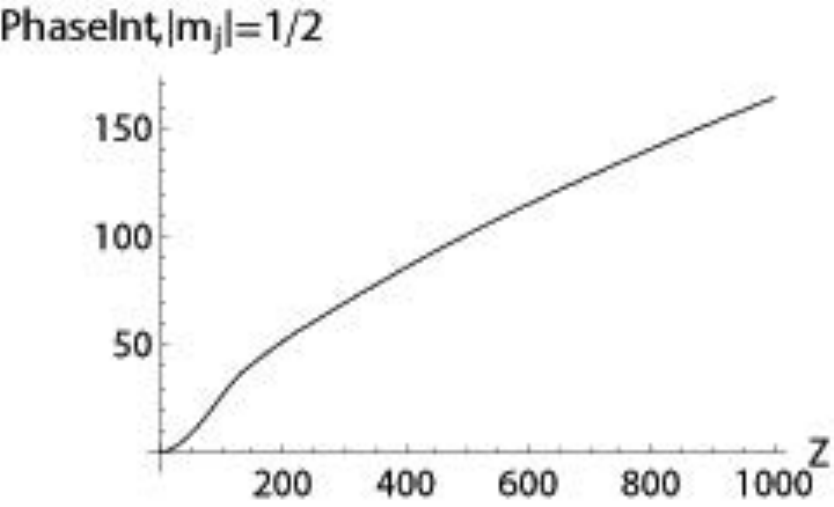} \\ a) }
	\end{minipage}
	\hfill
	\begin{minipage}{0.48\linewidth}
		\center{ \includegraphics[scale=0.7]{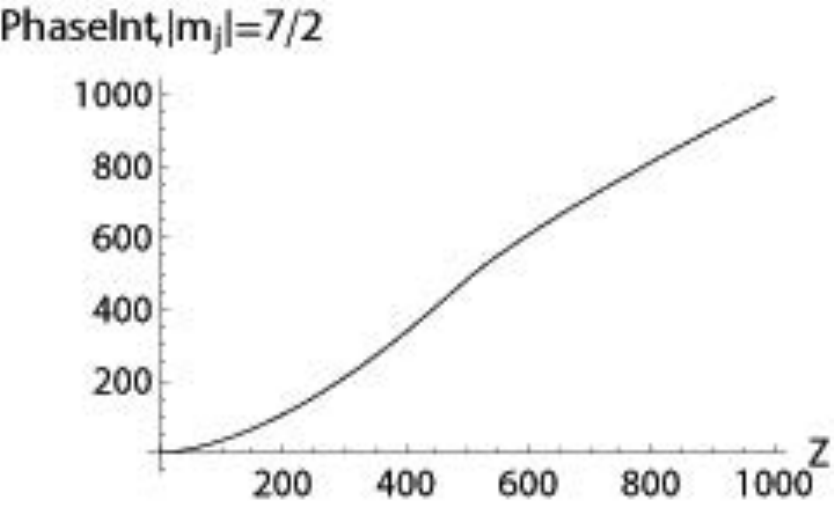} \\ b) }
	\end{minipage}
	\begin{center}
		{\small Fig.4. The dependence of the phase integral on $ Z $ for (a): $ |m_j|=\1/2 $; (b): $|m_j|={7 \over 2}$ .}
	\end{center}
\end{figure}

The typical behavior of the total bound energy of discrete levels (with account for the coefficient 1/2 in (\ref{f41}),(\ref{renenergy})) is shown in Fig.5. The total bound energy is a discontinuous function with jumps emerging each time, when the charge of the source reaches the corresponding critical value, the next discrete level dives into the lower continuum, and so the bound energy loses an amount of $ 2\, ( \times mc^2 )$. On the intervals between two neighboring  $ Z_{cr} $ the bound energy is always positive and increases monotonically, since there grow the bound energies of all the  discrete levels.  But unlike  1+1 D, in 2+1 D the total bound energy of discrete levels is always substantially less than the phase integral, what is clearly seen in the Figs.4,5.
\begin{figure}[h!]
	\begin{minipage}{0.48\linewidth}
		\center{ \includegraphics[scale=0.7]{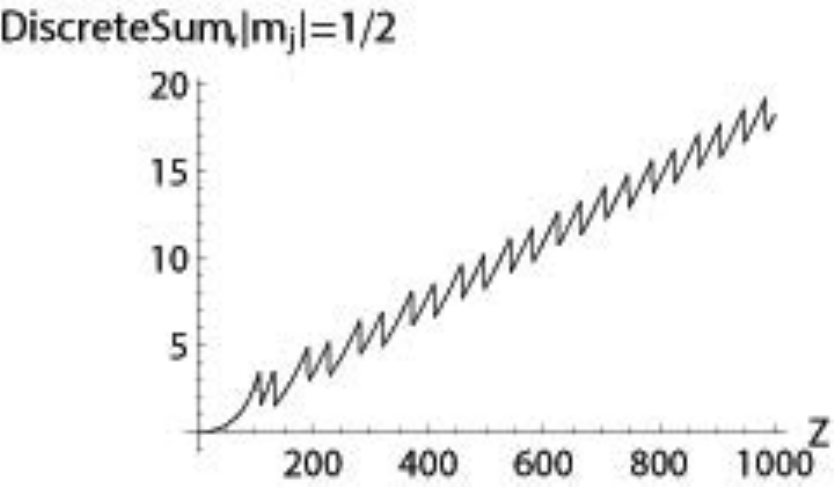} \\ a) }
	\end{minipage}
	\hfill
	\begin{minipage}{0.48\linewidth}
		\center{ \includegraphics[scale=0.7]{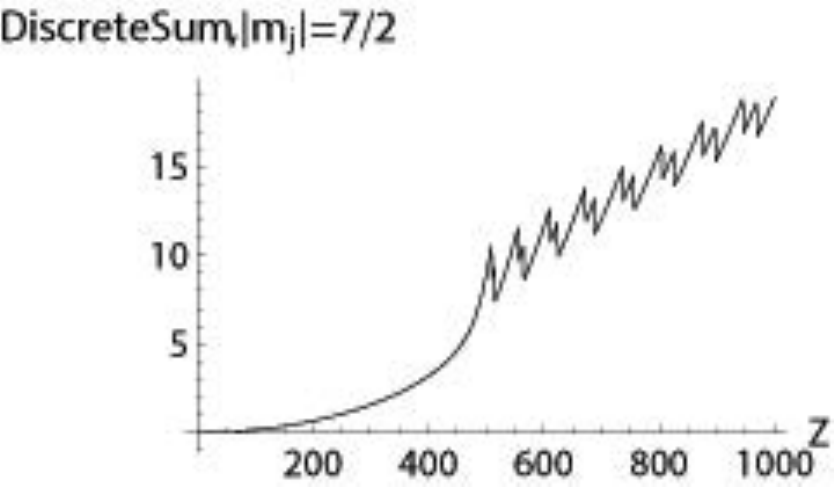} \\ b) }
	\end{minipage}
	\begin{center}
		{\small Fig.5. The dependence of the total bound energy of discrete spectrum on $ Z $ for (a): $ |m_j|=\1/2 $; (b): $|m_j|={7 \over 2}$ .}
	\end{center}
\end{figure}

 For more details of the whole picture the behavior of $ \E_{VP,|m_j|}^{ren}(Z) $ is shown in two versions. In the Fig. 6a,b there are presented separately the specific features of the partial channel with $|m_j|=1/2$, since in this channel the structure of the renormalization coefficient  $ \eta_{1/2} $ differs from the others with $|m_j|\not=1/2$. Namely, $ \eta_{1/2} =\eta_{PT}-\eta_{B,1/2}  $, where $\eta_{PT}$ corresponds to the PT-part of the renormalization coefficient, whereas $\eta_{B,1/2}  $ --- to the first term of the Born series for $\r_{VP,1/2}$ (see (\ref{f44}) and subsequent comments).
\begin{figure}[h!]
	\begin{minipage}{0.48\linewidth}
		\center{ \includegraphics[scale=0.7]{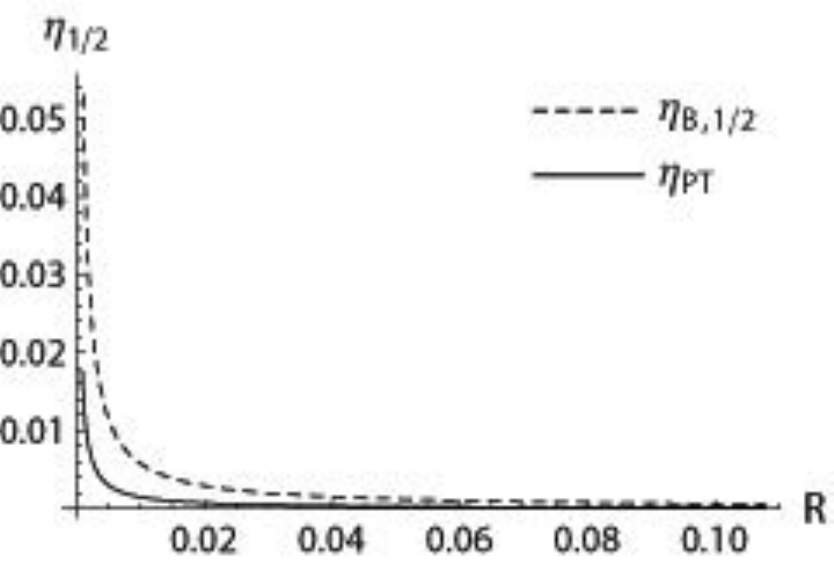} \\ a) }
	\end{minipage}
	\hfill
	\begin{minipage}{0.48\linewidth}
		\center{ \includegraphics[scale=0.7]{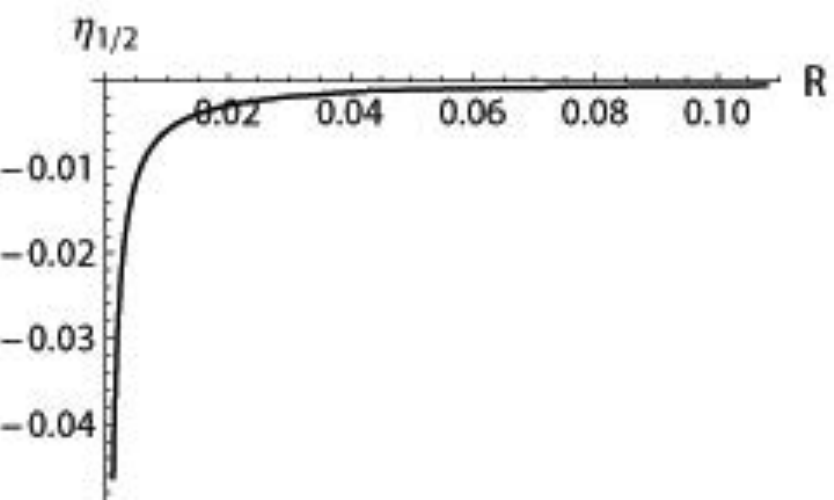} \\ b)  }
	\end{minipage}
	\begin{center}
{\small Fig.6a,b. The structure and details of behavior of the renormalization coefficient   $ \eta_{1/2} $ as a function of the radius $R$ of the central sphere.}
	\end{center}
\end{figure}

Furthermore, this channel demonstrates most clearly the change of the behavior of the renormalized vacuum energy from the perturbative quadratic growth for $Z \ll Z_{cr,1}$, when the dominant contribution comes from  $\E^{(1)}_{VP}$, obtained via PT according to (\ref{2.1})-(\ref{2.6}), to the regime of decrease into the negative region with increasing $Z$ beyond  $Z_{cr,1}$ (see Fig.7). Let us note that the Figs. 6,7  up to irrelevant details and concrete values of critical charges reproduce  the behavior of the renormalization coefficient and energy in the $s$-channel for the analogous three-dimensional problem.
\begin{center}
	\includegraphics[scale=1]{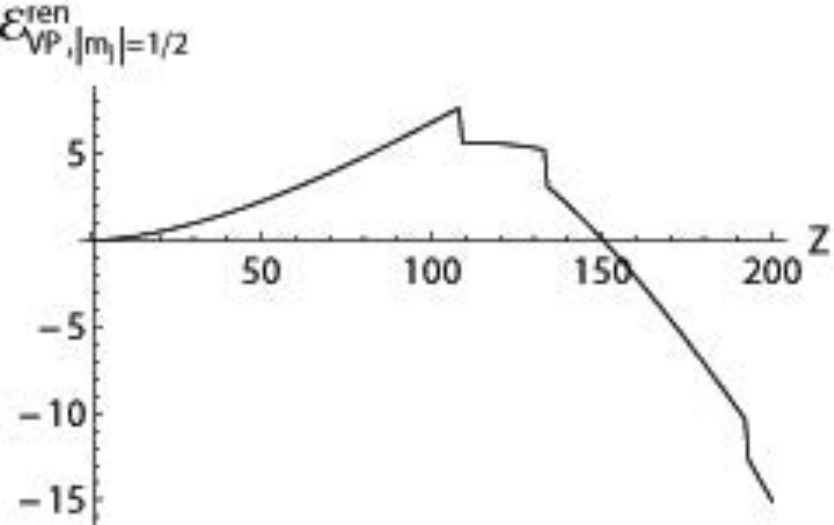} \\
	{\small Fig.7. The behavior of $ \E_{VP,1/2}^{ren}(Z) $ on the starting interval $0<Z<200$ with the change from the perturbative quadratic growth to the regime of decrease into the negative region by transition through $Z_{cr,1}$ and $Z_{cr,2}$.}
\end{center}

In the Fig.8 the dependence of $ \E_{VP,|m_j|}^{ren}(Z) $ on $Z$ in the interval $0< Z <1000$ for the three most representative values $ |m_j|=1/2\, , 5/2\, , 9/2 $, and also of the total renormalized energy $\E_{VP}^{ren}(Z) $, is shown.
\begin{figure}[h!]
	\begin{minipage}{0.48\linewidth}
		\center{ \includegraphics[scale=0.7]{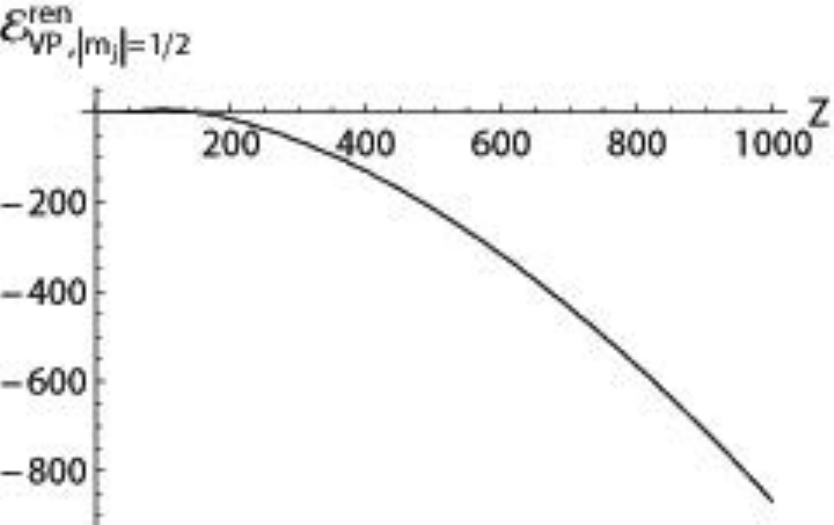} \\ a) }
	\end{minipage}
	\hfill
	\begin{minipage}{0.48\linewidth}
		\center{ \includegraphics[scale=0.7]{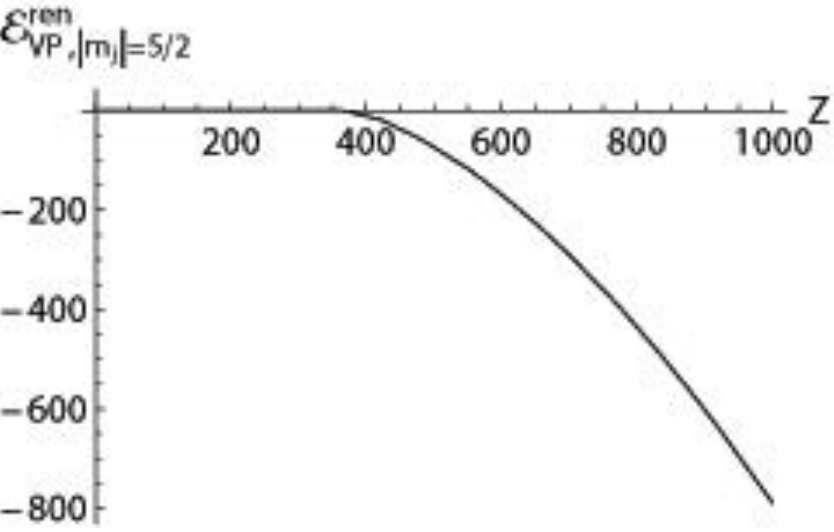} \\ b) }
	\end{minipage}
	\begin{minipage}{0.48\linewidth}
		\center{ \includegraphics[scale=0.7]{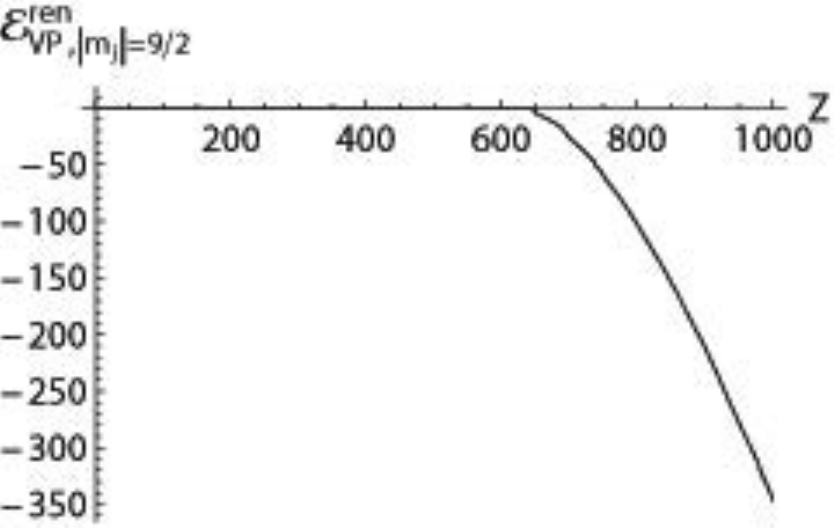} \\ c) }
	\end{minipage}
	\hfill
	\begin{minipage}{0.48\linewidth}
		\center{ \includegraphics[scale=0.7]{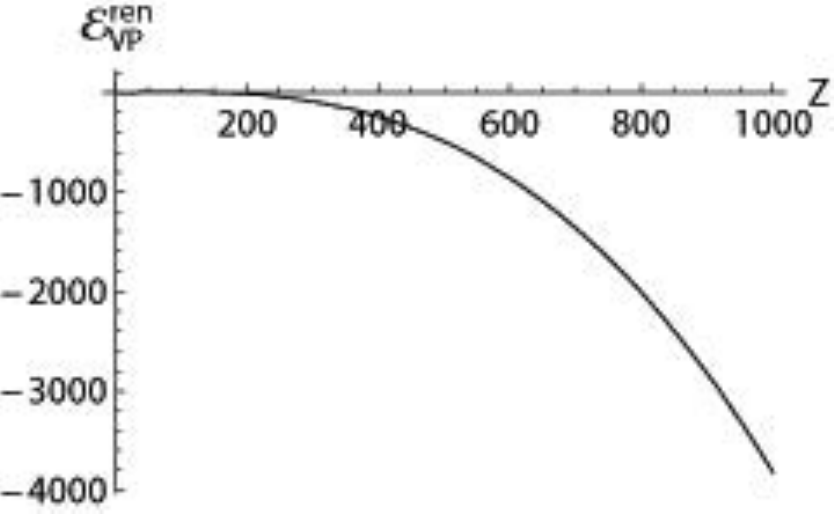} \\ d) }
	\end{minipage}
	\begin{center}
		{\small Fig.8(a-c): $ \E_{VP,m_j}^{ren}(Z)$  on the interval $0<Z<1000$ for $ |m_j|=\1/2 $ (a); $ |m_j|={5 \over 2}$ (b); $|m_j|={9 \over 2}$ (c). Fig.8d: the total renormalized energy $ \E_{VP}^{ren}(Z) $ on the same interval of $Z$}.
	\end{center}
\end{figure}
The histograms, demonstrating the contribution of different partial channels in $\E_{VP}^{ren}(Z) $, are presented in the Fig.9 for $Z=500$ and $Z=1000$.
\begin{figure}[h!]
\begin{minipage}{0.48\linewidth}	
\center{\includegraphics[scale=0.7]{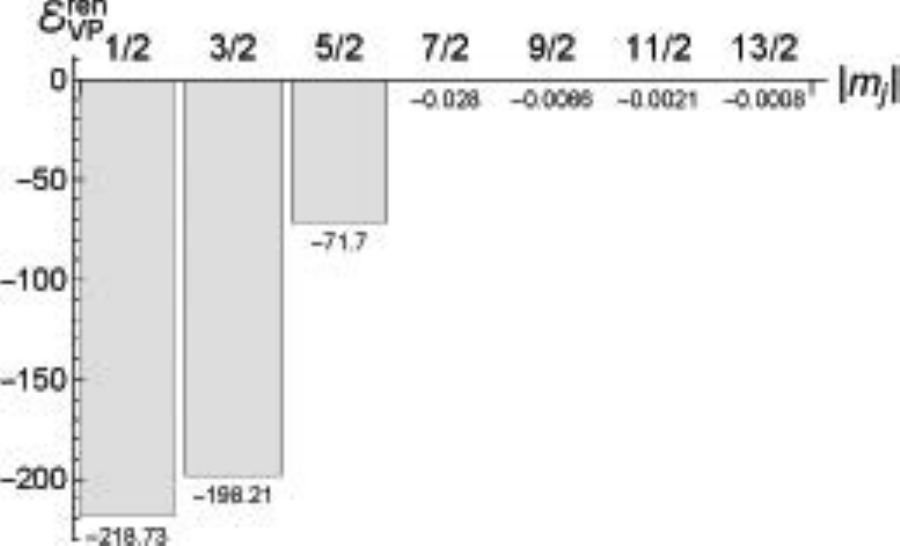} \\ a) }
\end{minipage}
\hfill
\begin{minipage}{0.48\linewidth}
\center{\includegraphics[scale=0.5]{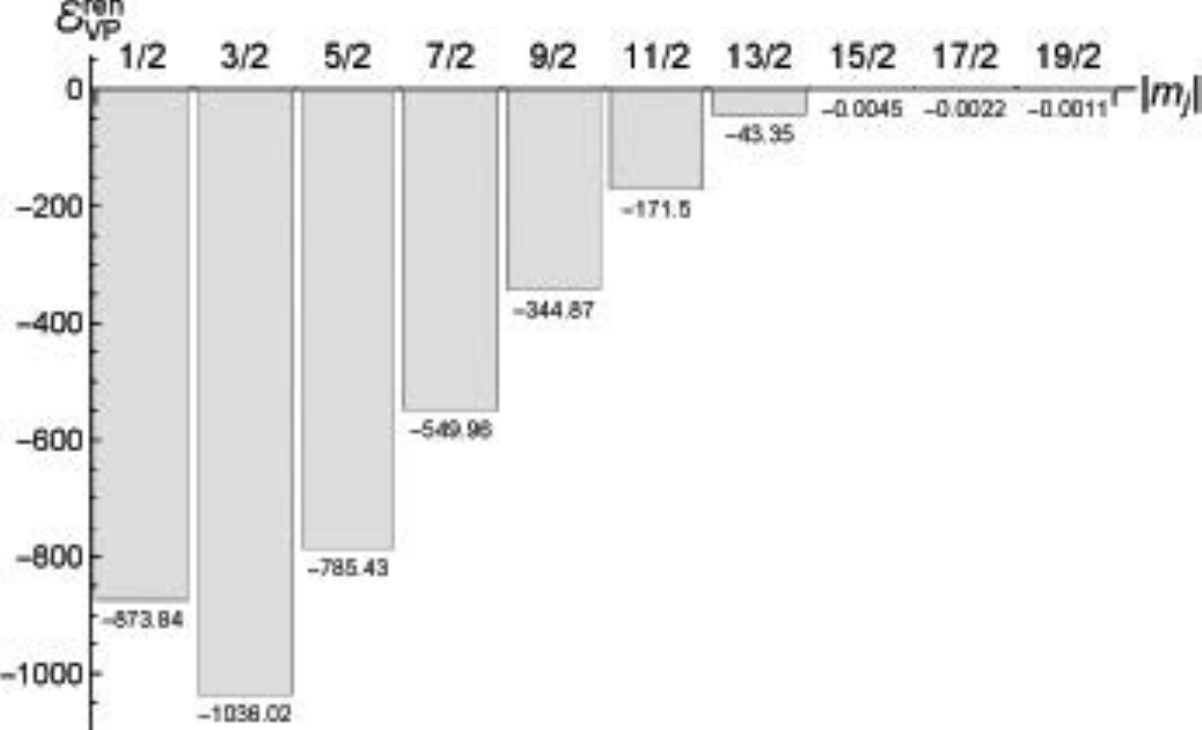} \\ b) }
\end{minipage}
\begin{center}
		{\small Fig.9. The contribution of different $ |m_j| $ to $\E_{VP}^{ren}(Z) $ for (a): $Z=500$; (b): $Z=1000$ \ .}
	\end{center}
\end{figure}	
In the Fig.10 for the same $Z$ the histograms, demonstrating the contribution from various partial channels to the total number $N(Z)$ of levels, that have reached the lower continuum, are shown. By comparing the Figs.9,10 it should be clear that the main contribution to the vacuum energy for the given $Z$ is produced indeed  by  those partial channels, where the discrete levels have already started to attain the lower continuum.
\begin{figure}[h!]
\begin{minipage}{0.48\linewidth}	
\center{\includegraphics[scale=0.65]{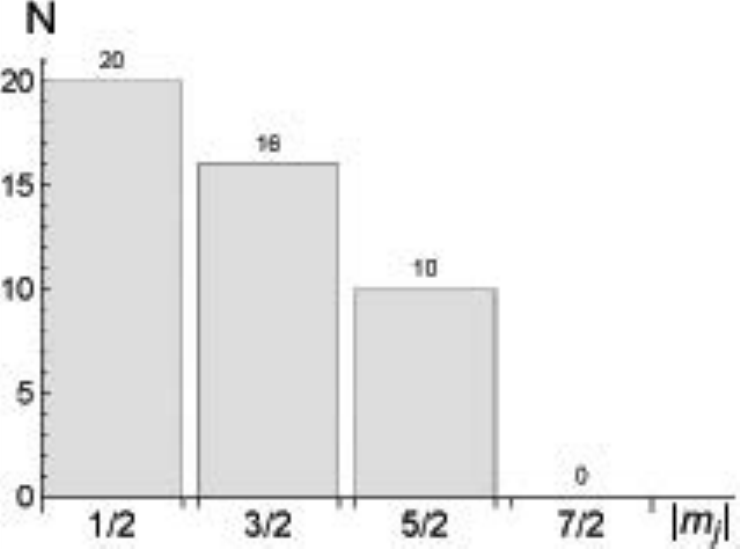} \\ a) }
\end{minipage}
\hfill
\begin{minipage}{0.48\linewidth}
\center{\includegraphics[scale=0.7]{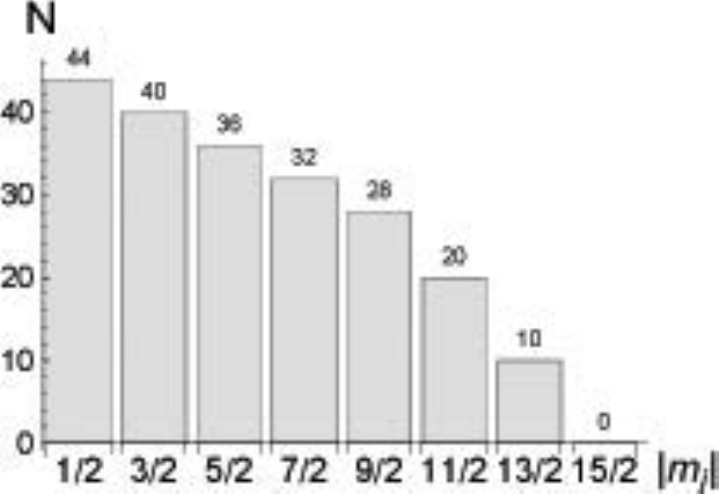} \\ b) }
\end{minipage}
\begin{center}
		{\small Fig.10. The contribution from various $ |m_j| $ to $Q_{VP}^{ren}(Z) $ for (a): $Z=500$; (b): $Z=1000$ \ .}
	\end{center}
\end{figure}	
In the Fig.11 the curves for $\E_{VP}^{ren}(Z) $ and for the total number $N(Z)$ of the vacuum shells, formed from the levels, which have sunk into the lower continuum, and for their power-like approximations are presented. A satisfactory approximation for $\E_{VP}^{ren}(Z) $ is given by the function  $ {\tilde \E}_{VP}(Z)=-1.55 \times 10^{-5} \times Z^{2.8} $, while the number of vacuum shells with sufficient accuracy is approximated by dependence ${\tilde N}(Z)=6.69 \times 10^{-5} \times Z^{2.17}$. It should be noted, however, that both these  approximations turn out to be just the estimates  for the true dependence of $\E_{VP}^{ren}(Z) $ and  $N(Z)$ on $Z$ for $Z \gg Z_{cr,1}$ from below, since only the range $0 < Z < 1000$, in which for  $0<Z<Z_{cr,1}$ both functions reveal completely different behavior, is used. By extending the range of approximation in $Z$ to the right the growth rate of $\E_{VP}^{ren}(Z) $ and  $N(Z)$  increases, but such  $Z$ by default lie beyond the scope of consideration in this work.
\begin{figure}[h!]
	\begin{minipage}{0.48\linewidth}
		\center{ \includegraphics[scale=0.7]{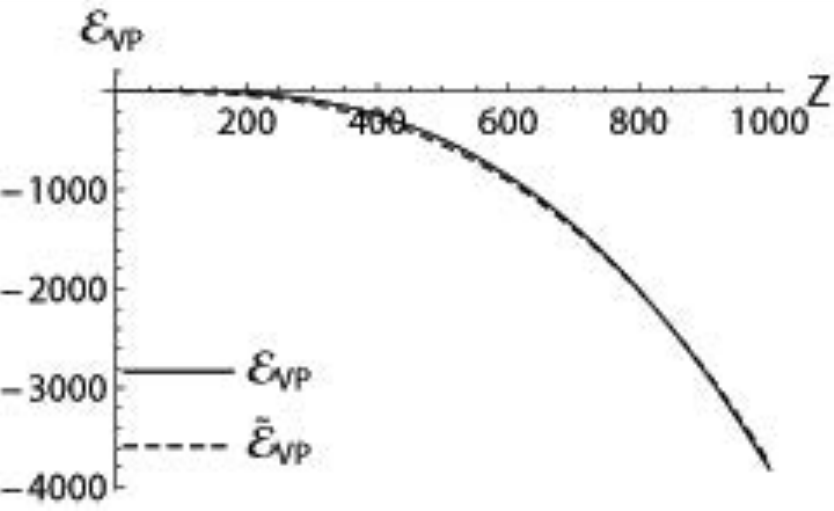}  }
	\end{minipage}
	\hfill
	\begin{minipage}{0.48\linewidth}
		\center{ \includegraphics[scale=0.7]{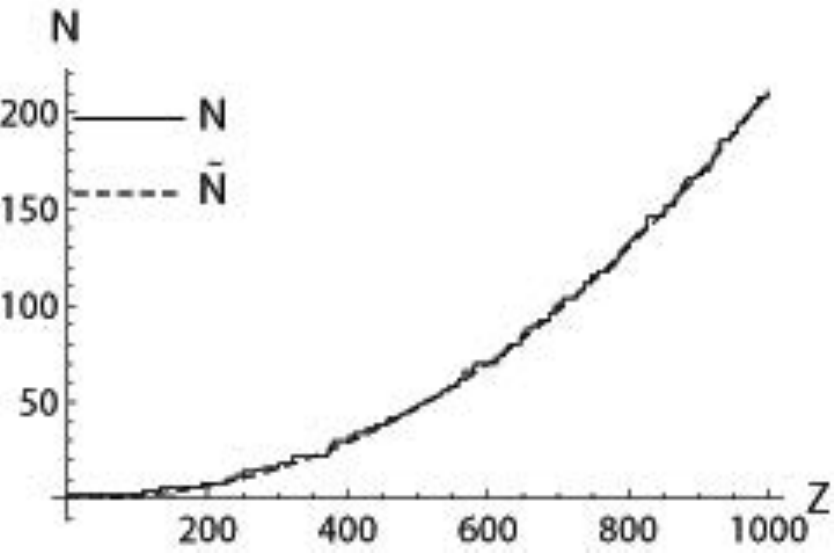}  }
	\end{minipage}
	\begin{center}
		{\small Fig.11. The vacuum polarization energy  $\E_{VP}^{ren}(Z) $, the number of vacuum shells $N(Z)$ and their approximations via power-like functions.}
	\end{center}
\end{figure}

Proceeding further, it would be worth noticing the following property of  $\E_{VP}^{ren}$. When for the fixed $Z \gg Z_{cr,1}$ the dependence $\E_{VP}^{ren}$ on the radius of the central sphere $R$ in the potential (\ref{1.00}) is considered, ignoring the relation (\ref{1.2}) between $R$ and  $Z$ any more,  then this dependence  by shifting $R$ from the reference point  $R_0(Z)=1.2\, (2.5\, Z)^{1/3}$ at least by one order in both directions turns out to be very close to $1/R$. So in the considered range of variation of  $Z$ and $R$ in the overcritical region $\E_{VP}^{ren}$   behaves like  $-\eta_{eff}\, Z^3/R\, , \ \eta_{eff}>0$, modulated by already sufficiently more slower function  like the logarithmic one. This conclusion follows from Figs. 12a-d, where  $\E_{VP}^{ren}$ and $1/\E_{VP}^{ren}$ as the functions of the ratio  $R/R_0(Z)$ for  $Z=1000$ are shown. It should be noted that by itself such a fast decrease  of $\E_{VP}^{ren}(Z) $ into the region of large negative values by decreasing ratio $R/R_0(Z)$ is easily explained through the increasing growth rate of the total number of discrete levels, diving into the lower continuum, thence through the growing rate of the vacuum shells number. Analogously, by increasing ratio  $R/R_0(Z)$  the levels should reach the threshold of the lower continuum more and more slowly, therefore $\E_{VP}^{ren}(Z) $ should increase. Moreover, for each $Z>Z_{cr,1}$ there  should always exist such  $R_{max}(Z)$, after reaching which the discrete levels could not attain the lower continuum at all, hence,  $\E_{VP}^{ren}(Z) $ becomes positive again. For $Z=1000$  $R_{max}(Z)$ lies in the range $R_{max}(Z)/R(Z) \simeq 70-75$, but since for $R$ close to $R_{max}(Z)$ the behavior of $\E_{VP}^{ren}$ becomes jump-like, while the behavior of $1/\E_{VP}^{ren}$  turns out to be almost completely irregular due to proximity of $\E_{VP}^{ren}$ to the zero mark, this range of ratio $R/R_0(Z)$ is not shown at all. It suffices to observe the distortion of the curve  $1/\E_{VP}^{ren}$ in the range $10<R/R_0(Z)<50$.
\begin{figure}[h!]
\begin{minipage}{0.48\linewidth}	
\center{\includegraphics[scale=0.7]{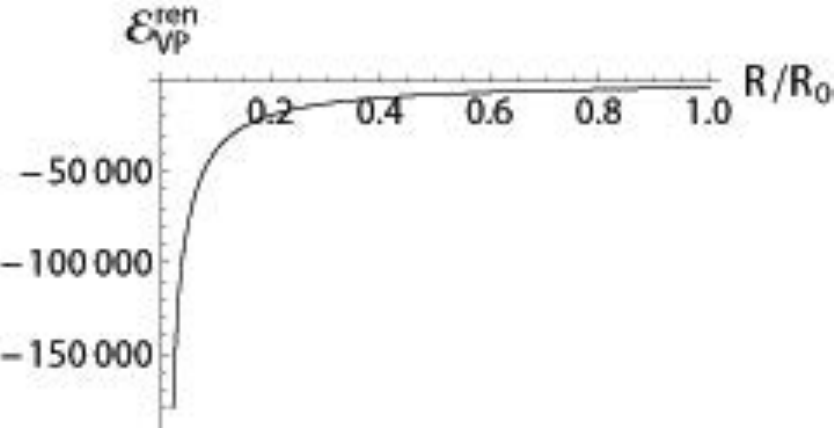} \\ a) }
\end{minipage}
\hfill
\begin{minipage}{0.48\linewidth}
\center{\includegraphics[scale=0.7]{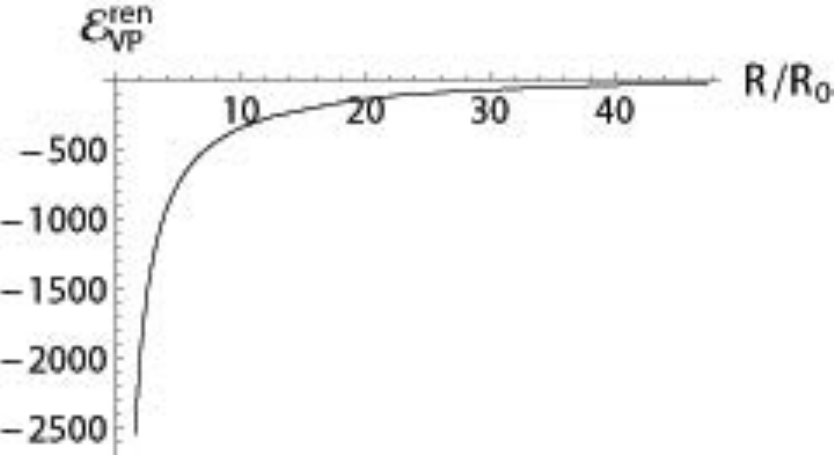} \\ b) }
\end{minipage}
\begin{center}
		{\small Fig.12a,b. $\E_{VP}^{ren}$ as a function of the ratio $R/R_0$ for $Z=1000$ (a): in the range $0< R/R_0 <1$; (b): in the range $1< R/R_0 <50$.}
	\end{center}
\end{figure}	
\begin{figure}[h!]
\begin{minipage}{0.48\linewidth}	
\center{\includegraphics[scale=0.7]{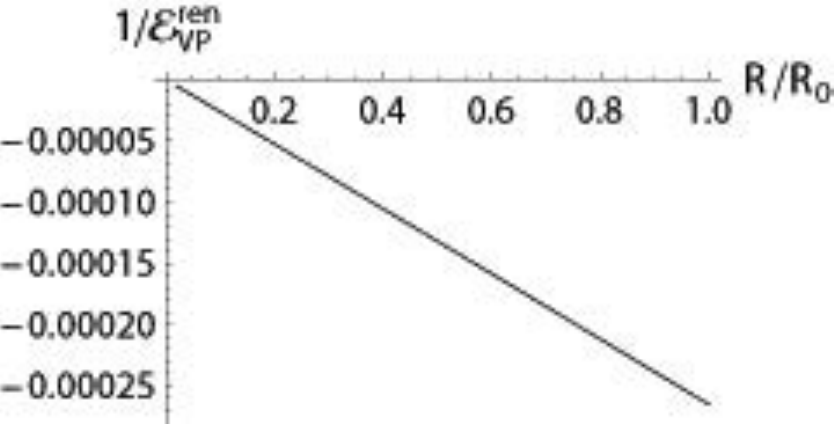} \\ c) }
\end{minipage}
\hfill
\begin{minipage}{0.48\linewidth}
\center{\includegraphics[scale=0.7]{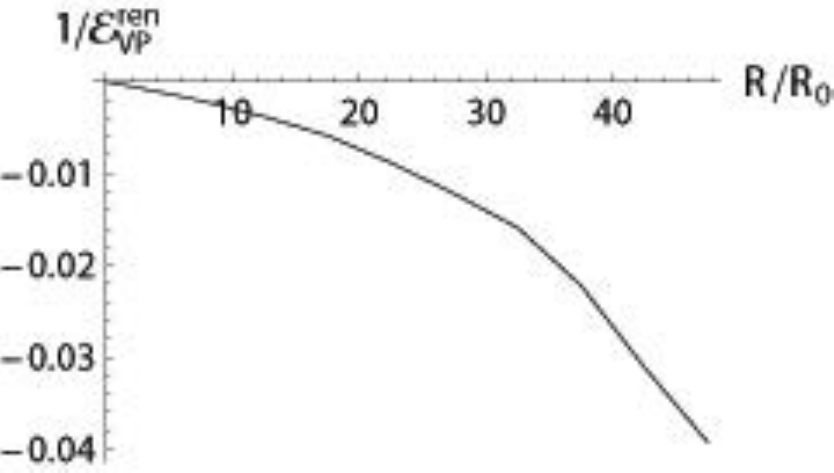} \\ d) }
\end{minipage}
\begin{center}
		{\small Fig.12c,d. $1/\E_{VP}^{ren}$ as a function of the ratio $R/R_0$ for $Z=1000$ (a): in the range $0< R/R_0 <1$; (d): in the range $1< R/R_0 <50$.}
	\end{center}
\end{figure}	

Let us also mention that the odd degree in the factor $Z^3/R$ in $\E_{VP}$ means that the case  $Z>0$ is considered. In essence,  $\E_{VP}$  is a definitely even function of $Z$, since it is connected with  $\r_{VP}$ via the well-known Schwinger relation  \cite{plunien1986,sveshnikov2017}
\beq
\label{f46}
\d \E_{VP}=\int \r_{VP} \d A_0^{ext} + \d \E_N  \ ,
\eeq
where  $\r_{VP}$ is always an odd function of  $Z$ by construction (see part I), while $\E_N$ is a discontinuous piecewise constant function, whose jumps appear each time when the next discrete level dives into the lower continuum and the bound energy loses $ 2\, (\times mc^2) $, and which also is an even function of $Z$ \cite{plunien1986,sveshnikov2017}. To underline this circumstance, in the general case the large Z-estimate for $\E_{VP}^{ren}$ should be written in the form $-\eta_{eff}\, |Z|^3/R\, , \ \eta_{eff}>0\, , $ with the cubic nonlinearity of the vacuum energy being the peculiar feature of vacuum polarization in 2+1 D.

\section{Conclusion}

To conclusion it should be  mentioned first of all that as in the case of the one-dimensional ``hydrogen atom'' ~\cite{davydov2017,sveshnikov2017,voronina2017}, actually in 2+1 D  the calculation of the vacuum energy by means of the UV-renormalization via fermionic loop could be implemented solely on the basis of relations  (\ref{f40}),(\ref{f41a}) and  (\ref{renenergy}) without applying to the vacuum density and shell effects. It is essential that by such renormalization we simultaneously ensure the convergence of the whole partial series for $\E_{VP}^{ren}$, since according to  (\ref{term_m}) the divergent terms in the sum (\ref{f41a}) are proportional to $(Z\a)^2$. So the renormalization via fermionic loop turns out to be the universal tool, that removes the divergence of the theory both in the purely perturbative and in the essentially non-perturbative regimes of vacuum polarization by the external Coulomb field.

From this point of view the effect of decreasing $\E_{VP}^{ren}(Z)$ in the overcritical region $\sim - \eta_{eff} Z^3$ could be explained quite correctly via the properties of the partial series (\ref{f41a}). Each separate term of this series $\E^{ren}_{VP,|m_j|}(Z)$ reveals the structure (\ref{renenergy}), which in essence is quite analogous to $\E_{VP}^{ren}(Z)$ in 1+1 D ~\cite{davydov2017,sveshnikov2017,voronina2017}. The direct consequence of the latter is that the negative contribution from the renormalization term $\eta_{|m_j|} Z^2$ turns out to be the dominant one in $\E^{ren}_{VP,|m_j|}(Z)$ in the overcritical region, since in this region  the growth rate of the non-renormalized energy in each separate channel, as in 1+1 D, should be $\sim Z^\n \ , \ 1<\n<2$. However, now the total number of the levels, which have sunk into the lower continuum  for the given  $Z$, is created by the sum of contributions from the finite number of first partial channels  with $|m_j|\leq |m_j|_{max}(Z)$, where  $|m_j|_{max}(Z)$ is the last one, in which the number of dived into the lower continuum discrete levels is non-zero. Indeed these channels yield the main contribution to the vacuum energy (see histograms in the  Figs.9,10). At the same time, for finite $Z$ $|m_j|_{max}(Z)$  is always finite  and grows approximately linearly with increasing $Z$.  And since the total vacuum energy   $\E_{VP}^{ren}(Z)$ is determined now mainly by the sum of contributions from  these channels, its rate of decrease  acquires an additional factor of order  $\mathrm{O}(Z)$, what in turn leads to the final answer  $\sim - \eta_{eff}\, Z^3$ in the overcritical region.

However, in essence the decrease of $\E_{VP}^{ren}$ in the overcritical region is governed first of all by the non-perturbative changes in the vacuum density for $Z > Z_{cr,1}$ due to discrete levels, reaching the threshold of the lower continuum (``the shell effect''). In 1+1 D the growth rate of the vacuum shells is $\sim Z^s \ , \ 1<s<2$, at least in the considered in Refs.~\cite{davydov2017}-\cite{voronina2017} range of external parameters. Therefore in the overcritical region the growth rate of the non-renormalized energy $\E_{VP}$ does not exceed $\sim Z^\n$, $1<\n<2$, and so the dominant contribution to $\E_{VP}^{ren}$ comes from the renormalization term $\eta Z^2$. In 2+1 D (and especially in 3+1 D) the shell effect is much more pronounced, the growth rate  of the total number of vacuum shells $N(Z)$ exceeds definitely  $\mathrm{O}(Z^2)$, and as a result $\E_{VP}^{ren}(Z)$ decreases in the overcritical region at least by one order of magnitude faster. Such behavior of  $\E_{VP}^{ren}(Z)$  in the overcritical region confirms once more the correct status of the assumption of the neutral vacuum transmutation into the charged one, which turns out to be the ground state of the electron-positron field in such external background \cite{raf2017,rg1977,greiner1985,plunien1986,greiner2012}, and hence of the spontaneous vacuum positron emission, which should accompany the creation of each subsequent vacuum shell due to the total  charge conservation.

It is worth-while noticing also that the methods of vacuum energy evaluation for the external potential of the type (\ref{1.00}), considered in this paper for 2+1 D, with minimal complements are carried over to the three-dimensional case. The main difference is that in 3+1 D there are two rotational quantum numbers --- $j m_j$, while the degeneracy of each energy level equals to $2j+1$ instead of 2. As a result, the partial series for the vacuum charge density and energy should reveal the structure  $2 \sum_\k\, |\k|f_{|\k|}$, where $\k=\pm(j+1/2)$, that in turn leads to increase of the growth rate of  $\E_{VP}^{ren}(Z) $ and   $N(Z)$ in the overcritical region by one more order compared to 2+1 D. More concretely, the performed calculations point out that with the same relation between the radius  $R$ and the charge $Z$ of the Coulomb source (\ref{1.2}) for $Z \gg Z_{cr,1}$ the total number of vacuum shells $N(Z)$ in 3+1 D behaves not slower than  $ \sim Z^{3.17}$, while the renormalized vacuum energy $\E_{VP}^{ren}(Z,R)$ decreases approximately as $-{\tilde \eta_{eff}}\, Z^4/R$.  In this case the rate of decrease  of the latter turns out to be so that it  becomes competitive with the classical electrostatic energy of the Coulomb source. The performed calculations show that in the case of the source in the form of a sphere the relation
 \beq
\label{f47}
\E_{VP}^{ren}(Z,R(Z))+ {Z^2 \a \over 2 R(Z)} \simeq 0  \ ,
\eeq
 will be satisfied at $Z^{\ast}\simeq 3000$. Thus, in  3+1 D the vacuum polarization effects turn out to be able for the total screening of the classical Coulomb reflection.  Moreover, the estimate $Z^{\ast}\simeq 3000$ turns out to a high degree a universal one, in particular, it depends very weakly on the deviation of the radius $R$ from the reference value  (\ref{1.2}) at least by one order of magnitude in both directions, and on the concrete structure of the Coulomb source as well. The Coulomb sources in the form of a uniformly charged sphere, charged ball, or a charged spherical layer lead to very close results for $Z^{\ast}\simeq 3000$. The main reason, why the detailed evaluation for the three-dimensional case is not published yet by the authors of the present paper, is that in the external Coulomb fields like (\ref{1.00}) the critical charge for the pion field amounts to $Z \sim 2000$ \cite{raf1978}, hence, for a consistent description of the vacuum polarization  in such fields the effects of strong interactions should be taken into account from the very beginning. Therefore the question, whether the other essentially nonlinear effects of vacuum polarization of the purely QED-origin could downgrade $Z^{\ast}$ up to values less than 2000, turns out to be quite actual. Without a consistent answer to this question it seems inappropriate for us to publish the details of the vacuum polarization energy evaluation in 3+1 D for such overcritical fields.

\bibliographystyle{apsrev4-1}
\bibliography{VP2D_II_arXiv}

\end{document}